\documentclass[useAMS,usenatbib]{mn2e}
\usepackage{graphicx}

\title[Optical and near-IR monitoring]{Optical and Near-IR long-term monitoring of NGC~3783 and MR~2251-178: evidence
  for variable near-IR emission from thin accretion discs.}
 
\author[P. Lira et al.]{P. Lira$^{1}$, P. Ar\'evalo$^{2,3}$, P. Uttley$^{4}$, I. McHardy$^{4}$ and E. Breedt$^{4,5}$\\
$^{1}$Departamento de Astronom\'{\i}a, Universidad de Chile, Casilla 36D, Santiago, Chile\\
$^{2}$Max-Planck-Institut f\"ur Astrophysik, Karl-Schwarzschild-Str.\ 1, D-85748 Garching, Germany\\
$^{3}$Departamento Ciencias F\'isicas, Universidad Andr\'es Bello, Av.\ Rep\'ublica 252, Santiago, Chile\\
$^{4}$School of Physics and Astronomy, University of Southampton, Southampton SO17 1BJ, UK\\
$^{5}$Department of Physics, University of Warwick, Coventry CV4 7AL, UK\\
}
\begin{document}

\date{}

\pagerange{\pageref{firstpage}--\pageref{lastpage}} \pubyear{2002}

\maketitle

\label{firstpage}

\begin{abstract}

  We present long term near-IR light curves for two nearby AGN: NGC~3783 and
  MR~2251-178. The near-IR data are complemented with optical photometry
  obtained over the same period of time. The light curves in all bands are
  highly variable and good correlations can be seen between optical and NIR
  variations. Cross-correlation analysis for NGC~3783 suggests that some disc
  near-IR emission is present in the J-band flux, while the H and K-bands are
  dominated by emission from a torus located at the dust sublimation
  radius. For MR~2251-178 the cross-correlation analysis and the
  optical--near-IR flux-flux plots suggest that the near-IR flux is dominated
  by disc emission.  We model the optical to near-IR Spectral Energy
  Distributions (SED) of both sources and find that disc flaring might be a
  necessary modification to the geometry of a thin disc in order to explain
  the observations. The SED of MR~2251-178 gives some indications for the
  presence of NIR emission from a torus. Finally, we consider the implications
  of the standard alpha disc model to explain the different origin of the
  variable NIR emission in these AGN.

\end{abstract}
 

\section{Introduction}
 
Variability was very early established as a trade mark of Active Galactic
Nuclei (AGN). Variability has also been used as a key tool to derive physical
properties of AGN: characteristic time scales were used to infer sizes of the
emitting regions, lags between the ionising continuum and the line response
have been used to determine Black Hole masses ($M$), multi-wavelength light
curves have been used to study the geometrical and physical connections
between the different regions around the central engine.

The scenario accepted until recently for the interplay of the emitting
regions is that variability is driven by the emission from the X-ray
corona located close to the central Black Hole (e.g., Collin-Souffrin,
1991; Krolik et al.\ 1991; Clavel et al., 1992; Collier et al.\ 1999;
Cackett et al.\ 2007). The negligible optical inter-band lags was
early evidence that pointed towards reprocessing of high-energy
photons by the accretion disc, where the characteristic distances
between the different emitting regions correspond to the light
travelling time (Collin-Souffrin, 1991; Krolik et al.\ 1991; Clavel et
al., 1992). Later on, the measurement of short lags between the X-ray
emission and the optical, and the leading of the X-rays whenever
significant lags were determined (e.g., Edelson et al., 1996; Wanders
et al., 1997; Shemmer et al., 2001; Desroches et al., 2006), lent
support to this picture (see also Nandra et al., 2000 for evidence
supporting reprocessing from the correlation between the UV flux and
the spectral shape of the X-ray emission), since for intrinsic disc
variability shorter wavelengths should lag longer wavelength emission
by long (viscous) time-scales. Also, light curves showed that the
amplitude of the X-ray variations was much larger than that seen in
the optical, which can be explained by the damping of the signal
during the disc reprocessing. However, a full picture of the interplay
between the X-ray corona and the disc might not be complete just yet,
as new evidence seems to suggest that long term optical variability is
driven by accretion rate fluctuations in the disc itself (see Section
3.1 and also Ar\'evalo et al., 2008, 2009, and references therein).

Long term ($\sim$ years), well sampled optical, UV, and X-ray
monitoring of AGN is available for several sources (e.g., Breedt et
al., 2010; Chatterjee et al., 2009; Shemmer et al., 2001; Giveon, et
al., 1999; T\"urler et al., 1999; Korista et al., 1995; Clavel et al.,
1991, Clavel, Wamsteker \& Glass, 1989). However, near-IR data of
similar quality has been lacking until recently. The new
implementation of robotic observations and queue-based operations of
other telescopes is changing this situation rapidly.

Suganuma et al.\ (2006) presented high quality optical and near-IR
light curves for 4 Seyfert 1 galaxies obtained over 3 years of intense
monitoring. The analysis of these observations, together with some
other data available in the literature, showed that the lags observed
at near-IR wavelengths are in very good agreement with the location of
a putative dusty torus whose inner face should be found at the dust
sublimation radius, i.e., at a distance proportional to $L^{1/2}$,
where $L$ corresponds to the luminosity of the central source. This
behaviour is expected for a torus that intercepts a significant
fraction of the near-UV and optical flux from the central source and
reradiates it in the near and mid-IR. In this scenario, any flux
variations in the central source will be followed by variations at
these near-IR wavelengths, with a delay that will be characterised by
the location of the torus inner face, which corresponds to the
distance at which dust no longer sublimates under the strong glare of
the central radiation.

Near-IR emission, however, can be emitted by the colder regions of an
accretion disc itself, although there is no clear theoretical
predictions of the real extent of the outer regions of the
disc. Kishimoto et al.\ (2008) have presented evidence for near-IR
disc emission in Quasars in polarised flux. They find that the
spectral energy distribution is very close to $f_{\nu} \propto
\nu^{1/3}$, as expected for a Shakura-Sunyaev $\alpha$ disc. More
recently Landt et al.\ (2011) have shown that emission from an
accretion disc and hot dust are required to explain the continuum
around 1 micron in a sample of AGN.

In this paper we will present near-IR observations for NGC~3783 and
MR~2251-178, and argue that we have detected near-IR variability from their
accretion discs. This paper is organised as follows: In Section 2 we
characterise our targets; in Section 3 we describe the acquisition and
analysis of the data; in Section 4 we give estimates for the host contribution
to our nuclear photometry; in Section 5 we present our results in the form of
light curves, cross correlation analysis, flux-flux plots and spectral energy
distributions; finally, Section 6 discusses our findings and the summary is
presented in Section 7. In this article we adopt a concordance $\Lambda$CDM
cosmology with $H_0$ = 70 km/s/Mpc.

\section{NGC~3783 and MR~2251-178 Black Hole Masses}

NGC~3783 is a well studied nearby Seyfert galaxy located at a distance of 42
Mpc ($z=0.0097$). Early reverberation mapping campaigns were reported by
Reichter et al.\ (1994) and Stirpe et al.\ (1994). Revised results were later
reported by Onken \& Peterson (2002) and Peterson et al.\ (2004), finding a
black hole mass of $M = 2.98 \pm 0.54 \times10^7 M_{\odot}$.

For MR~2251-178, a very nearby Quasar ($z=0.0640$), a black hole mass
of $M \sim 2 \times10^8 M_{\odot}$ was recently reported by Wang, Mao
\& Wei (2009) using spectroscopic data and applying the
luminosity-radius relation for the Broad Line Region as calibrated by
Green \& Ho (2005). We have also obtained spectroscopic data for
MR~2251-178 using the RC spectrograph on the 1.5m telescope at CTIO
and operated by the SMARTS consortium. The data were obtained on the
20th of December 2006 and reduced and calibrated in the standard way.
 
Following Peterson et al.\ (2004) we measure a line-width for H$\beta$
of 4,145 km/s. Using our light curves (presented in Section 5.1) we
determine an average 5100\AA\ flux of $\sim 9\times10^{-15}$
ergs/s/cm$^{2}$/\AA. For a distance of 274 Mpc, we obtain a black hole
mass of $M \sim 2.0 \pm 0.5 \times10^8 M_{\odot}$, in complete
agreement with the value determined by Wang, Mao \& Wei (2009). These
estimates are also in agreement with the lack of an observed break in
the X-ray Power Spectrum Density, which puts an lower limit to the
black hole mass of $> 10^8 M_{\odot}$ (Summons et al., in
preparation).

\section{Data}
 
\subsection{Optical Observations and previous results}

Optical and X-ray monitoring for NGC~3783 and MR~2251-178 were already
presented elsewhere (Ar\'evalo et al., 2008; Ar\'evalo et al.,
2009). The analysis of the optical and X-ray light curves showed that
while the time delay observed between the X-ray and the optical bands
for MR~2251-178 was consistent with $0\pm4$ days, NGC~3783 required
that some optical flux was produced about 6 light days away from the
region where the X-rays were produced (Ar\'evalo et al., 2008, 2009).

Crucially, both sources showed that optical long term variations (time
scales of years) presented very large fluctuations which could not be
explained by the reprocessing of the X-ray light curve. Hence,
Ar\'evalo et al.\ (2008, 2009) have suggested that the long term
variability seen in the optical is driven by the intrinsic
fluctuations of the accretion flow and is characterised by the disc
viscous time scales.

In this work we have extended the coverage of the optical light curves
by one year for NGC~3783 and two years for MR~2251-178. The data
initial reduction followed the same steps detailed in Ar\'evalo et
al.\ (2008, 2009). The photometry was obtained taking all images to a
common seeing and then performing aperture photometry with a diameter
of 3.7 arc-seconds.

\subsection{Near-IR SMARTS observations}
 
The J, H, and K observations were obtained between the 1st of December 2006
and the 1st of August 2009 for NGC~3783, and between the 1st of August 2006
and the 20th of November 2009 for MR~2251-178. We used the ANDICAM camera
mounted on the 1.3m telescope at CTIO and operated by the SMARTS
consortium. The field of view corresponds to 140 arc-seconds, with a pixel
size of 0.274 arc-seconds. Exposure times were 40 seconds in the J band and 20
seconds in the H and K bands. The mean seeing was $\sim 1.0-1.2\arcsec$, with
the smaller values found in the K-band. The average sampling of the light
curves was $\sim 4.5$ days with a period of intensive ($\sim$ daily)
monitoring during the second year of observations for both
sources. Simultaneous optical observations in the B and V-bands were also
obtained and reduced as outlined above.

The data reduction followed the standard steps of dark subtraction,
flat fielding, and sky subtraction using consecutive jittered
frames. The light curves were constructed from relative photometry
obtained through a fixed aperture of 2.7 and 3.6 arc-seconds in
diameter for NGC\,3783 and MR\,2251-178, respectively, after all
images were taken to a common seeing by convolving each image with a
Gaussian of width $=\sqrt(\sigma_0^2-\sigma^2)$, where $\sigma_0$ is
the width corresponding to some of the worst seeing conditions (dates
with even poorest seeing were discarded) and $\sigma$ is the width of
each individual image.  For NGC~3783 five different comparison stars
were used to obtain the relative photometry whenever
available. MR~2251-178 however, had only one comparison star within
the field of view.

Photometric errors were computed as the squared sum of the standard
deviation due to the Poissonian noise of the source and sky flux
within the aperture, plus the uncertainty of obtaining the measurement
itself. This last error was estimated as the dispersion in the
photometry of the stars available in the field of view between two
consecutive exposures and it was found to range from 0.02 to 0.05 in
fractional flux.

 \begin{figure*}
 \centering
 \includegraphics[scale=0.45,angle=0]{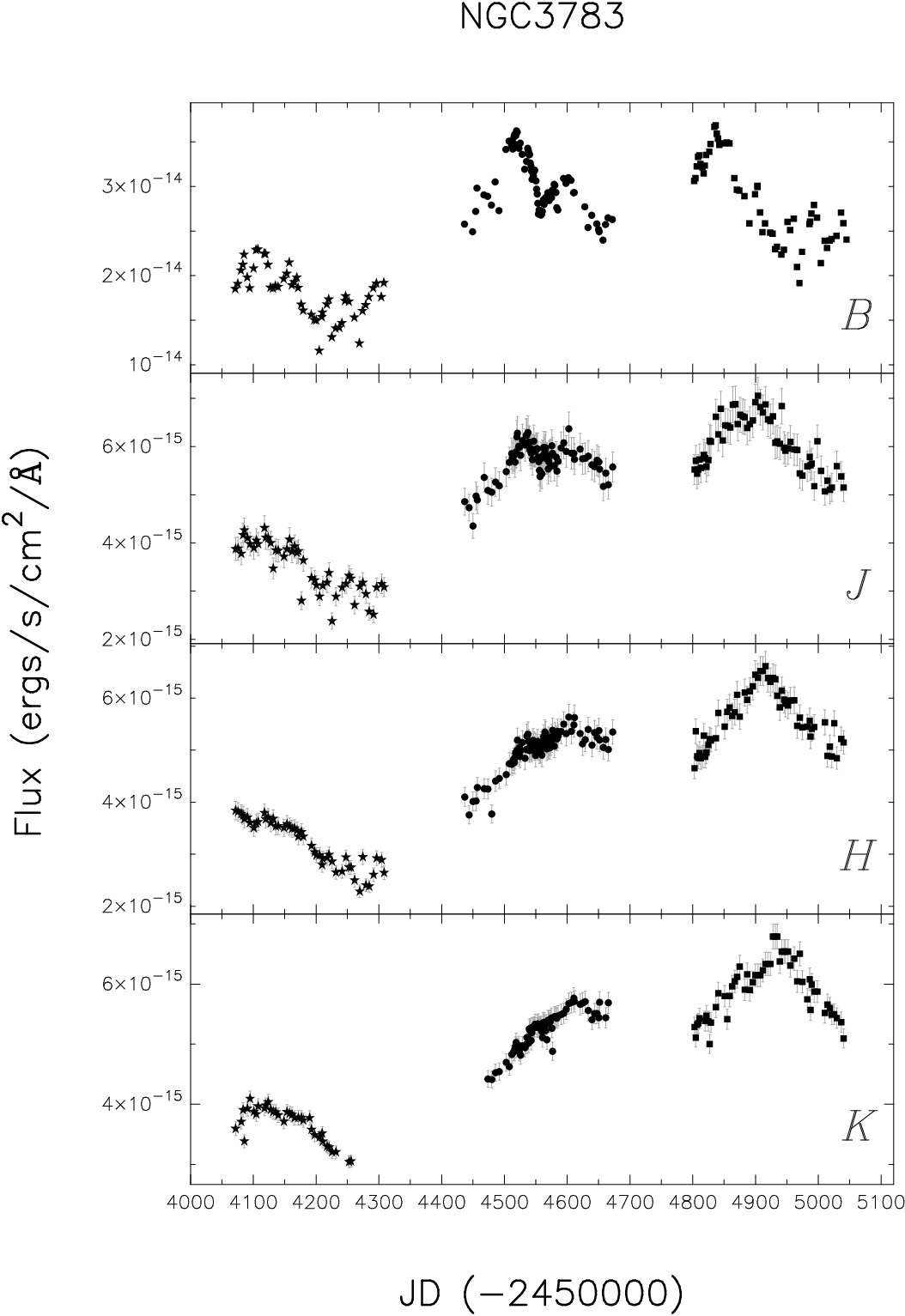}
 \hspace{0.4cm}
 \includegraphics[scale=0.45,angle=0]{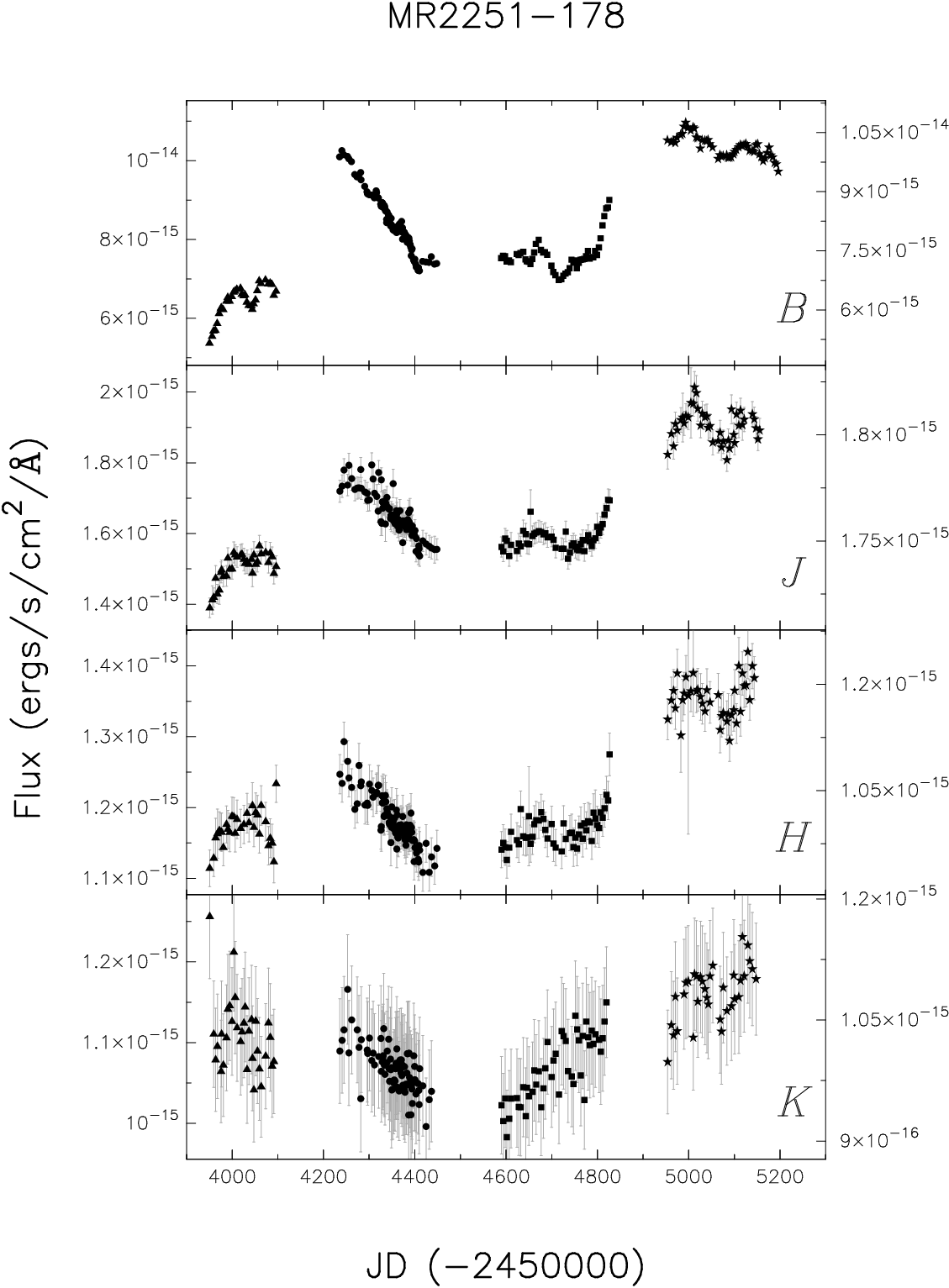}
 \caption{Optical and near-IR light curves for NGC~3783 {\bf (left)} and
   MR~2251-178 {\bf (right)}. Different segments (years) are shown with
   different symbols. For MR~2251-178 we show two scales to the y-axis,
   corresponding to a correction from host galaxy flux contribution to our
   photometry of 15\% (on the left) and 45\% (on the right) of the total bulge
   flux.}
 \end{figure*}

\section{Host Galaxy Contribution}

One component that should be taken into account when studying AGN
variability is the light contribution within the used aperture from
the stellar population of the host galaxy. High mass sources might be
spared from this correction if they have high accretion rates, but for
systems with massive black holes and low accretion rates it is now
known that the host will also contain a massive spheroid which should
be accounted for since the emission from the active nuclei might not
completely dominate the total observed flux. In the following
subsections we will estimate this contribution for NGC~3783 and
MR~2251-178.

\subsection{NGC~3783}

NGC~3783 was studied using the ACS camera on board the Hubble Space
Telescope by Bentz et al.\ (2006) and the galaxy was modelled as a
combination of a bulge, disc and bar in the F550M filter ($\sim
5100$\AA). The amount of host light in our 3.7\arcsec\ optical
apertures at this wavelength corresponds to $1.9 \times 10^{-15}$
ergs/s/cm$^2$/\AA, while for the 2.7\arcsec\ near-IR apertures it
corresponds to $1.5 \times 10^{-15}$ ergs/s/cm$^2$/\AA\ (M.\ Bentz,
private communication), with more than 95\% of this flux coming from
the bulge. To extrapolate this measurement at 5100\AA\ to near-IR
wavelengths we will assume that only the bulge contributes
significantly to the measured fluxes and will use the spectral energy
distribution of the appropriate stellar population.

The stellar populations in bulges vary as a function of their
luminosity. Massive bulges are characterised by old stellar populations which
are well represented by a single epoch of star formation (Peletier et al.,
1999; Peletier \& Balcells, 1997), while the colours of less massive bulges
found in late type galaxies are better characterised when younger stellar
populations are also included (Carollo et al., 2007). From the determination
of the black hole mass in NGC~3783 the mass of the bulge is found to be $\sim
2\times 10^{10} M_{\odot}$ (H\"aring \& Rix, 2004), at the lower limit of the
range in the Peletier's sample, and therefore it should be well represented by
a stellar population with 1 to 2 times solar metallicities and ages around
9-12 Gyr. 

We have used the simple stellar population models from the work of
Maraston et al.\ (2005), scaled to the HST measurement at 5100\AA, to
determine the host contributions to the near-IR fluxes. A correction
is introduced to account for the different apertures used during the
photometric measurements in the optical and near-IR images. It is
found that, after scaling, the single stellar population model with
$Z=1 Z_{\odot}$ and 9 Gyr contributes less to the near-IR flux than
the model with $Z=2 Z_{\odot}$ and 12 Gyr by a factor of 1.4--1.5,
while the contributions are nearly equal in the B and V bands. The
final B-band value reported in Table 1 corresponds to the average from
these two extreme values and agrees within 25\%\ with the value
derived by Alloin et al.\ (1995) using a similar aperture. For a
visualisation of the contributions, see Section 5.4.1.

\begin{table} 
\centering
\caption{Host contribution within a 2.7\arcsec aperture in diameter. All
  fluxes are in units of ergs/s/cm$^2$/\AA. For NGC~3783 the fluxes represent
  the average of single stellar populations taken from Maraston et
  al.\ (2005). For MR~2251-178 $BJHK$-band total bulge flux estimates come
  from Marconi \& Hunt (2006). The $V$-band estimation comes from the adoption
  of the same stellar populations as for NGC~3783. For MR~2251-178 corrections
  corresponding to a 15\% and 45\% of the total bulge flux are given.}
\begin{tabular}{cccccc} \hline
Galaxy & $f_B$ & $f_V$ & $f_J$ & $f_H$ & $f_K$ \\ \hline
NGC~3783    & 1.51e-15 & 2.20e-15 & 1.18e-15 & 7.89e-16 & 3.43e-16 \\
MR~2251-178 & 1.0e-16 & 1.7e-16 & 1.2e-16 & 8.7e-17 & 3.9e-17 \\ 
MR~2251-178 & 3.2e-16 & 5.0e-16 & 3.7e-15 & 2.6e-16 & 1.2e-16 \\ 
\hline
\end{tabular}
\end{table}

\subsection{MR~2251-178}

For MR~2251-178 there is no study of the host galaxy, except for
H$\alpha$ imaging of the very extended ionised nebula around the
quasar (see Shopbell, Veilleux \& Bland-Hawthorn, 1999, and references
therein). We have therefore estimated the host contribution to our
photometry in an indirect way. Marconi \& Hunt (2003) have determined
a tight correlation between the black hole mass and the total bulge
luminosity in the B and near-IR bands which can allow us to infer the
bulge light contribution to our photometry.

Using the estimate of the black hole mass in MR~2251-178 we can
determine the total bulge near-IR flux. Next it is necessary to
interpolate to the flux within our aperture. To do this we need to
assume a characteristic bulge effective radius ($r_e$) and Sersic
index ($n$). Marconi \& Hunt's sample suggests that for a $\sim 10^8\
M_{\odot}$ black hole $r_e \sim 3-5$ kpc is appropriate (although
values of $r_e$ as low as 1 kpc and as high as 9 kpc are also being
observed for this mass range). The work of Graham \& Driver (2007)
shows that $n \sim 3-4$ should be adopted. Computing the fraction of
flux within our apertures, it is found that this represents about
$\sim 30\%$ of the bulge light. The scatter in $r_e$, however, is
large, implying that the correction can range from 15\% to 45\% of the
total bulge flux.

This exercise is clearly a circular argument since we have not taken
into account the host contribution to our photometry when determining
the black hole mass in Section 2. However, while the dependency of the
luminosity-radius relation on the optical luminosity goes as
$L^{0.65}$, the dependence on the H$\beta$ width goes as FWHM$^2$
(e.g., Netzer \& Trakhtenbrot, 2007). Therefore our estimate is fairly
robust and correcting for the host contribution only lowers our
estimate of the mass by a small factor which is well within the errors
involved in the method.

The host contribution fluxes in our apertures are presented in Table 1 where
two values are shown for MR~2251-178 corresponding to contributions within our
aperture of 15\% and 45\% of the total bulge luminosities. Given the very
similar apertures used to obtain the optical and near-IR photometry for
MR\,2251-178, no aperture correction was applied in this case.

\section{Results}
 
\begin{table} 
\caption{Light curve characterisation. The characterisation of MR~2251-178 was
  obtained assuming a host contribution of 30\% of the total bulge determined
  following Marconi \& Hunt (2003); see Section 4.2. $<\!f\!>$ corresponds to
  the average flux, $\sigma$ to the standard deviation, $F_{var}$ is the
  normalised variability amplitude ($\sqrt(\sigma^2\! -
  \!\delta^2)/\!<\!f\!>$, where $\delta$ are the photometric uncertainties),
  and $R_{max}$ is the ratio of the maximum to minimum flux. For definitions
  see Suganuma et al.\ (2006). $<\!f\!>$ and $\sigma$ are given in units of
  ergs/s/cm$^2$/\AA.}
\begin{tabular}{cccccc} \hline
Galaxy & Band & $<\!f\!>$ & $\sigma$ & $F_{var}$ & $R_{max}$\\ \hline
NGC~3783    & $B$ & 2.48e-14& 6.19e-15& 2.92& 3.18\\
            & $J$ & 4.99e-15& 1.21e-15& 2.61& 2.96\\
            & $H$ & 4.60e-15& 1.13e-15& 2.72& 2.89\\\smallskip
            & $K$ & 5.06e-15& 9.87e-16& 2.00& 2.23\\ 
MR~2251-178 & $B$ & 8.48e-15& 1.41e-15& 2.14& 1.75\\ 
            & $J$ & 1.56e-15& 1.45e-16& 1.14& 1.37\\
            & $H$ & 1.13e-15& 8.26e-17& 0.88& 1.30\\
            & $K$ & 1.07e-15& 5.32e-17& 0.58& 1.26\\ \hline
\end{tabular}
\end{table}

 \begin{figure*}
 \centering
 \includegraphics[scale=0.3,angle=-90]{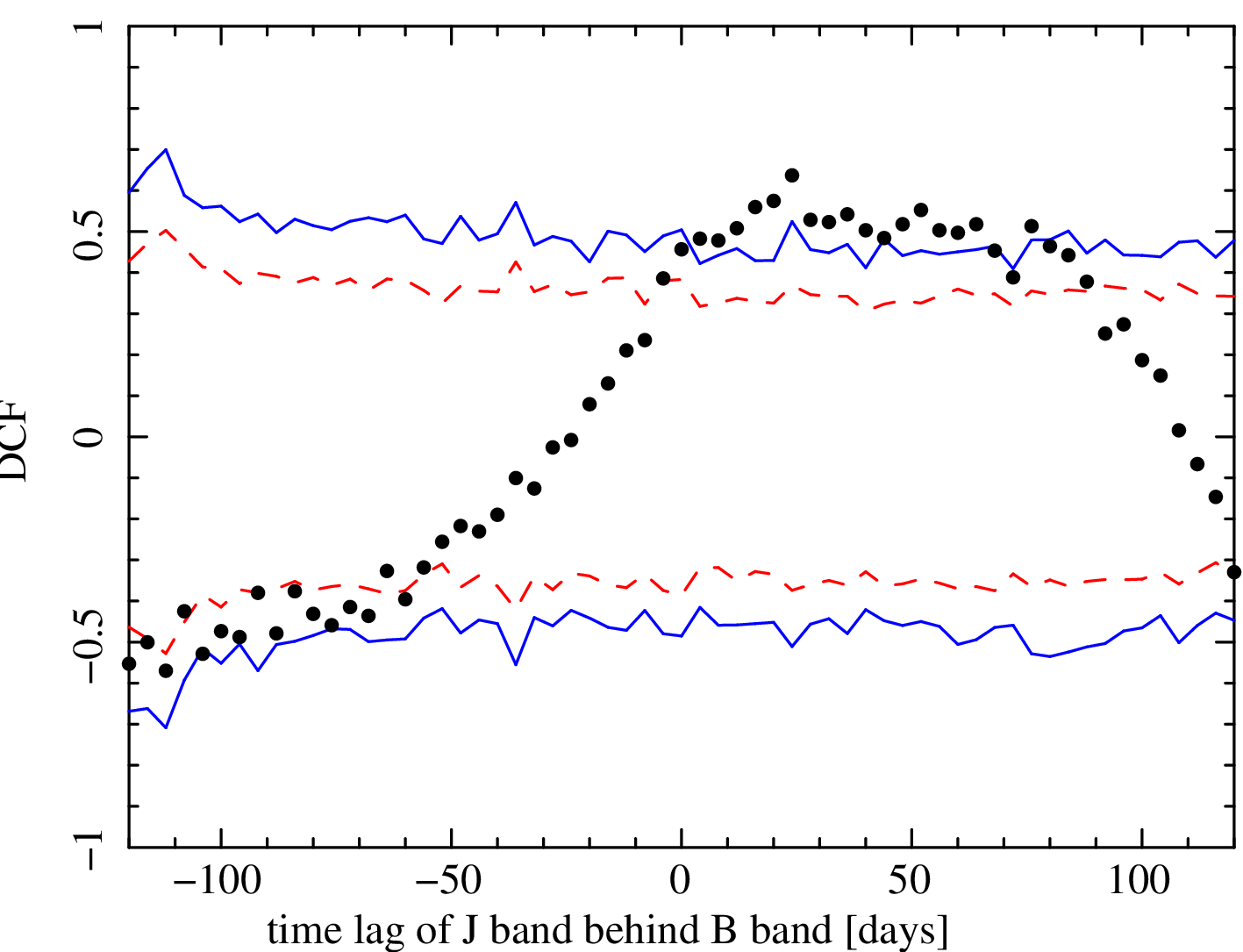}%
 \hspace{0.4cm}
 \includegraphics[scale=0.3,angle=-90]{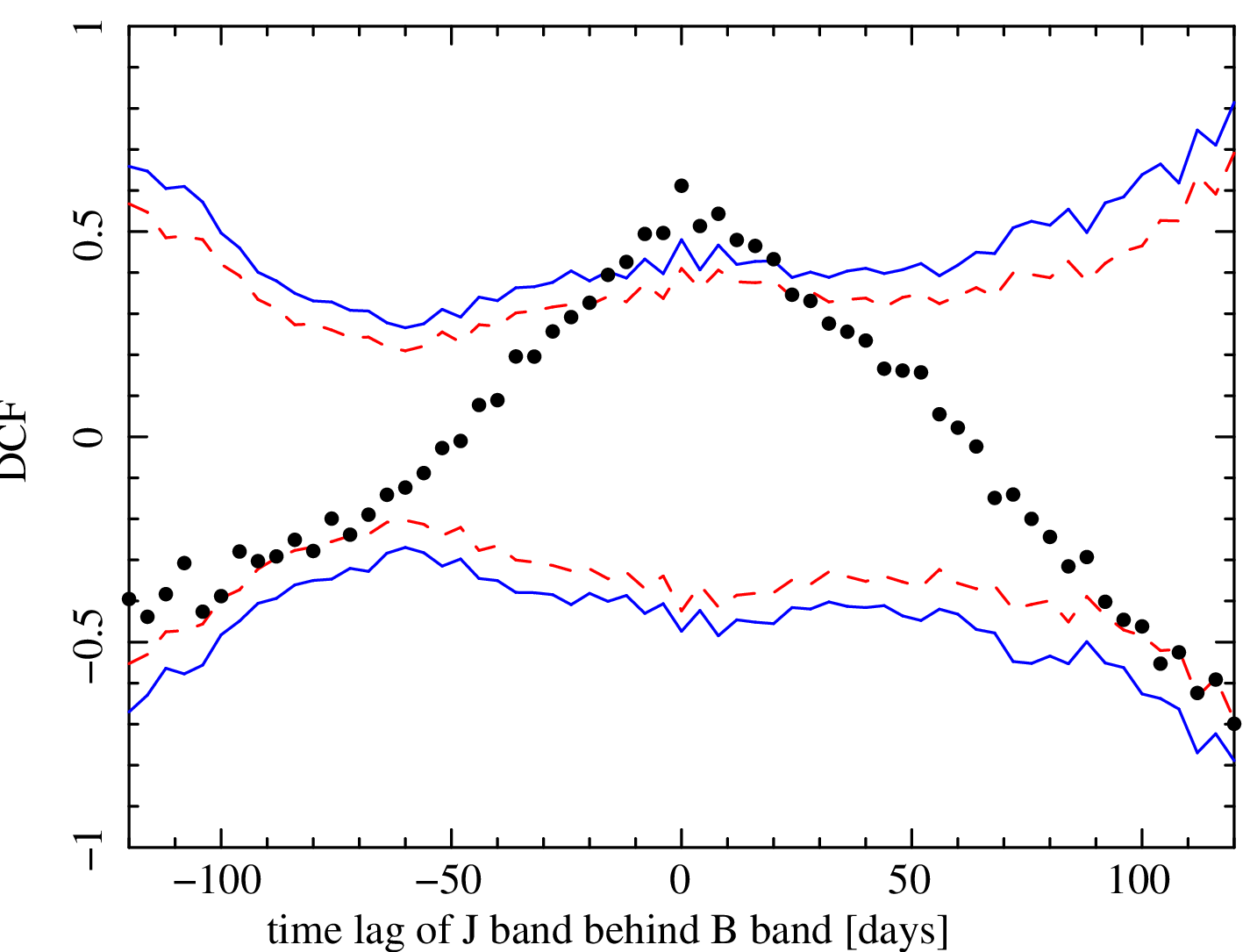}\\
 \includegraphics[scale=0.3,angle=-90]{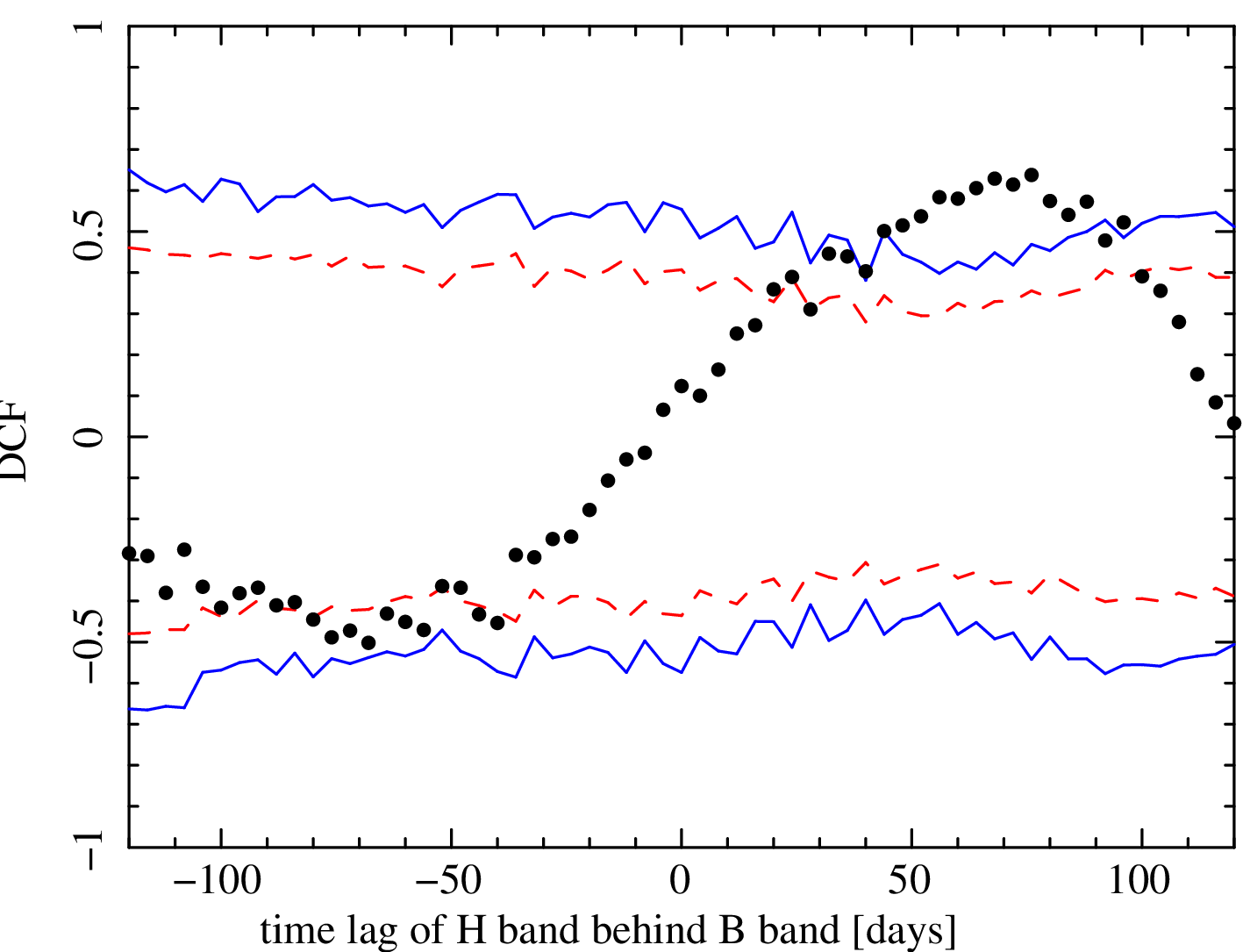}%
 \hspace{0.4cm}
 \includegraphics[scale=0.3,angle=-90]{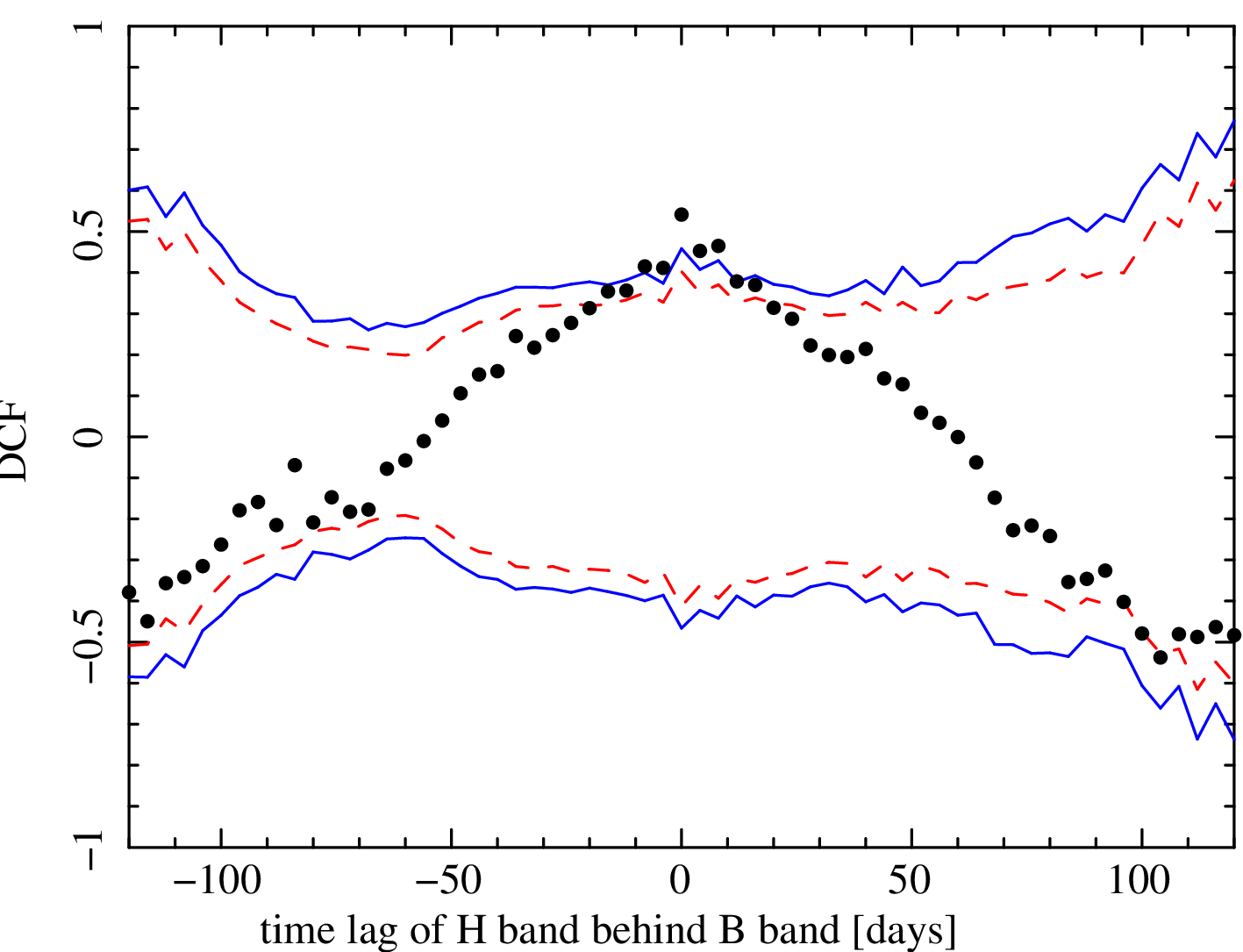}\\
 \includegraphics[scale=0.3,angle=-90]{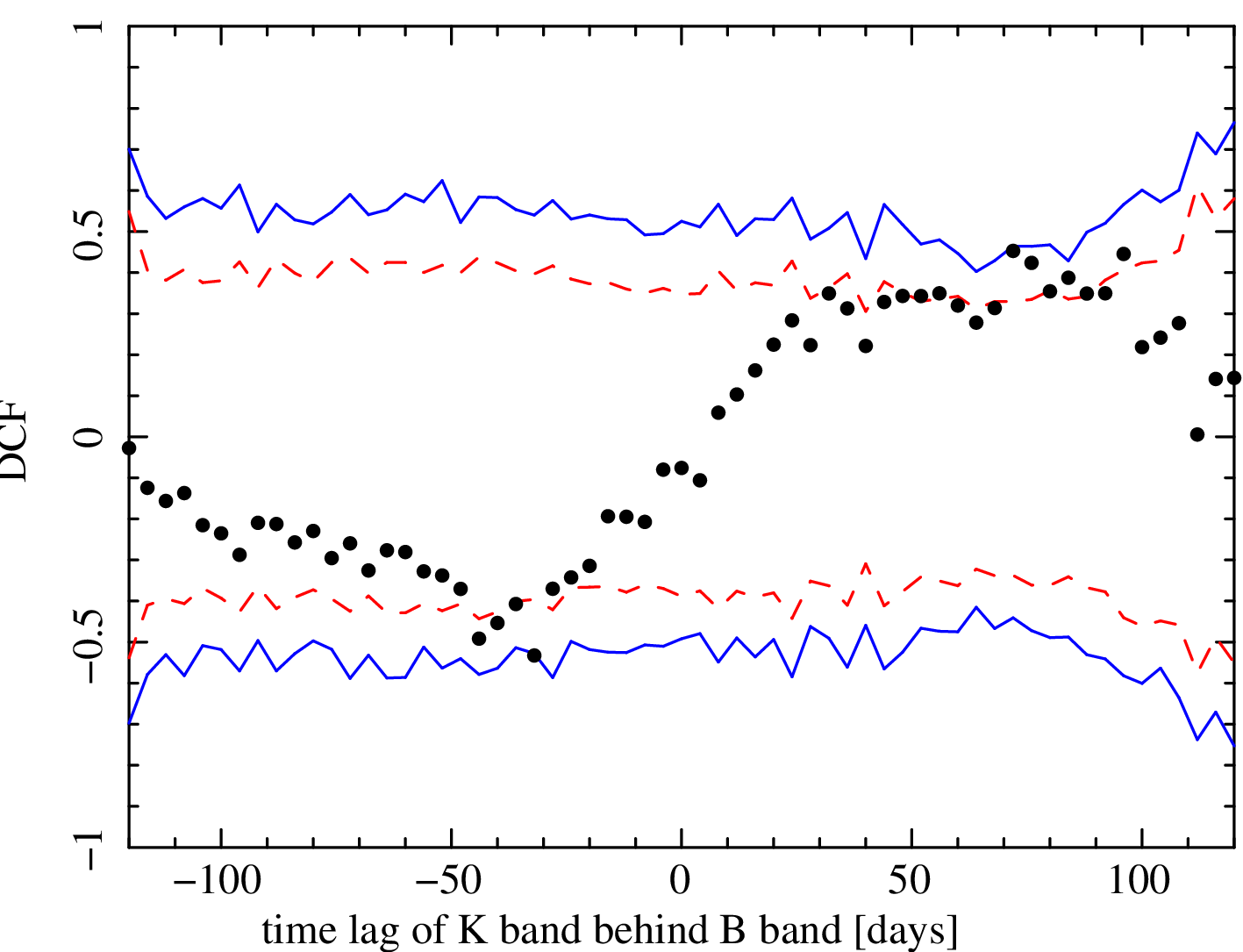}%
 \hspace{0.4cm}
 \includegraphics[scale=0.3,angle=-90]{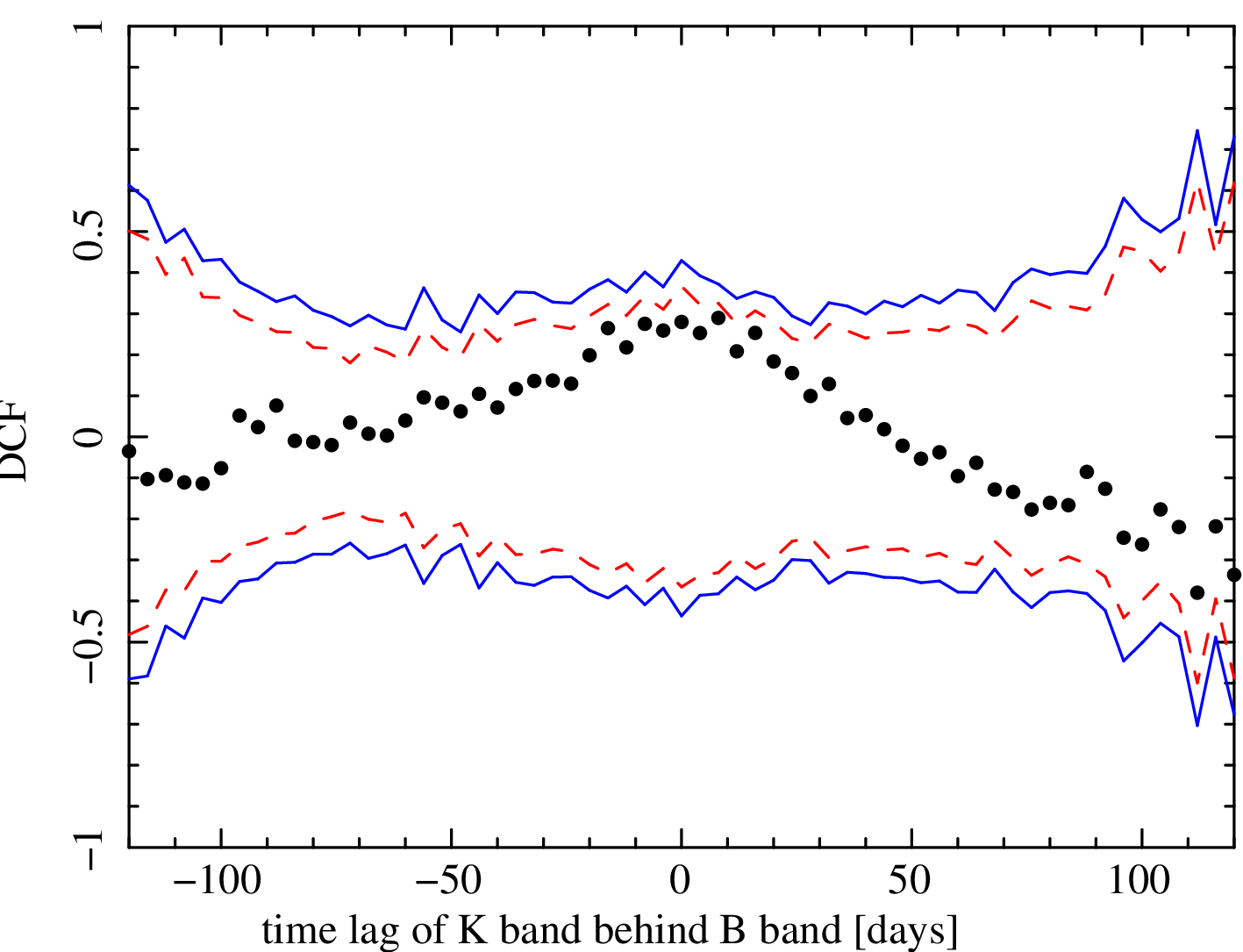}
 \caption{NGC~3783 {\bf(left)} and MR~2251-178 {\bf(right)} Discrete
   Correlation Functions between the B vs J bands (top), B vs H bands
   (middle), and B vs K bands (bottom). Positive lags mean that the near-IR
   bands lag behind the optical. The dashed and continuous lines represent 95\%
   and 99\% confidence limits as obtained from Monte Carlo simulations (see
   Ar\'evalo et al.\ (2008) for details).}
 \end{figure*}

\subsection{Light Curves}

The near-IR light curves are presented in Figure 1, together with the
optical B-band. All data have been corrected for foreground extinction
in the Milky Way and host contribution. Since this correction is not
certain for MR~2251-178 we show two scales to the y-axis, one
corresponding to a 15\% correction (on the left) and one corresponding
to a 45\% correction (on the right).

The light curve for NGC~3783 presents a very interesting behaviour,
where optical long term variations (time scales of months to years)
are well reproduced in the near-IR, but with short time variations
(time scales of days to weeks) being more obvious in the shorter
near-IR bands and possibly completely gone in the K-band. This can be
better seen in the first two years of monitoring where the J-band
follows closely the short time variability of the B-band light curve,
albeit with a smaller amplitude. The third year of monitoring clearly
shows that on longer time scales the near-IR bands lag behind the
optical light curve, with the lag being larger at longer wavelengths.

On the other hand, MR~2251-178 presents a good agreement between the
optical and all the near-IR bands. Interestingly, the fourth year of
monitoring presents an unusual pattern, where the fast variability
seems to be more significant in the J and H-bands than in the
optical. Unfortunately, the photometric errors are quite large in the
K-band, and the fast variation of the lightcurves cannot be assessed
in the same detail.

In Table 2 we present the most relevant characteristics for the light curves
after resampling the periods of intensive monitoring to the same cadence as in
the rest of the light curves. Notice the clear reduction in the amount of
variability towards longer wavelengths seen in both sources. 

\subsection{Cross Correlation Analysis}

To examine the relation between the regions responsible for the emission seen
in the optical and near-IR in a more quantitative manner, we obtained the
cross correlation between the light curves presented in the previous Section.

To estimate the cross correlation between the light curves, we used the
discrete correlation function (DCF) method of Edelson \& Krolik (1988). The
lags corresponding to the main peak in these functions, and their errors, were
estimated using the random sample selection method of Peterson et al.\ (2004),
selecting 67 percent of the data points in the B and NIR light curves and
calculating the DCF centroid for 1000 such trials. The lags quoted correspond
to the median of the distribution of trial centroids and the errors are the
67\% bounds. Centroids are calculated as weighted average of the DCF points
above 50\% of the peak value DCF$_{\rm max}$. Further details of the procedure
can be found in Ar\'evalo et al.\ (2008).

Figure 2 presents the optical--near-IR cross correlation obtained for
NGC~37783 and MR~2251-178. It can be seen that for MR~2251-178 the
cross correlation between the optical and the J and H bands presents a
strong peak consistent with a zero lag between the different
wavelengths. On the other hand, it was not possible to determine a
centroid for the K-band with a confidence level above 95\%. This is
likely due to a combination of two factors: the large error bars in
the light curve, and the small amplitude of the variation in this
band.  Notice, for example, that between the first and second year of
monitoring the B-band light curve presented a jump flux $\ga$ 50\%,
while the K-band flux level did not present any significant
increase. Despite this, the general behaviour of the K-band cross
correlation seems to follow the trend observed at shorter wavelengths.

However, the cross correlation of the optical and near-IR data for
NGC~3783 shows a very different behaviour, as expected from the light
curves already discussed. The cross correlation between the J and
B-band show a very wide peak, which extends from $\sim$ 10 days
(location where the correlation rises above the 99\% probability), to
$\sim$ 60--80 days (where the correlation finally falls below the 99\%
probability). In contrast, the correlation between the B and H-band
correlation shows a much narrower peak with a clear centroid around
60--70 days. This last finding is in very good agreement with the
results found by Glass (1992) who determined a lag of about 80 days
after 15 years of monitoring with a total of about 65 photometric
measurements in the U and K bands and interpreted it as the response
from a dusty torus. Glass (2004) reported a longer lag ($\sim 190$
days) but pointed out that the sampling of the new data were not
good. He also reported an average delay between the J and L-bands of
148 days, where the L-band presumably corresponds to emission arising
in colder regions of the dusty torus.

We interpret the behaviour of the cross correlation of the optical and
near-IR light curves in NGC~3783 as being due to the presence of two
emitting regions. While one region seems to dominate at shorter
wavelengths (primarily in the J-band) and follows closely the short
term variability of the optical light curves with a lag of a few days,
the second region seems to be much further away and essentially
follows the long term variations of the dusty torus. Using the
relation between $M_{V}$ and the distance at which the inner face of
the torus is located (Suganuma et al., 2006), we predict a distance of
$\sim 60$ light days for the dusty torus in NGC~3783. This is in very
good agreement with our findings. 

Beckert et al.\ (2008) determined the size of the inner face of the
dusty torus in NGC~3783 using MIDI VLTI observations and find a radius
of 0.2--0.3 pc, which corresponds to a light travel time of $\sim 300$
light-days (see also Kishimoto et al., 2009). This is however, a model
dependent determination and values much smaller are also allowed
(although not favoured) by Beckert et al.\ (2008). Besides, while the
mid-IR interferometric observations report on a light-weighted-size of
the emitting torus, near-IR variability studies are much more
sensitive to the innermost regions of the torus, where the response to
the variations of the central source are maximised. What has been
learnt from the mid-IR interferometric data, however, is that the
dusty structure surrounding the central source is quite extended
(Kishimoto et al., 2009).

Table 3 summarises the cross correlation analysis between the optical and
near-IR bands for NGC~3783 and MR~2251-178.

\begin{table} 
\caption{Cross Correlation Analysis. All lags are expressed in days and
  correspond to the mean centroids. A positive lag between bands $Y\!/Z$ means
  that $Y$ leads. An estimate of the Torus sublimation radius ($R_{T}$, in
  light days), based on the work of Suganuma et al.\ (2006), has also been
  added in the last column.}
\begin{tabular}{lccccc} \hline
Galaxy & $X/B$ lag & $B/J$ lag & $B/H$ lag & $B/K$ lag & $R_{T}$\\ \hline
NGC~3783     & $6.6^{+7.2}_{-6.0}$ & $40.8^{+5.1}_{-10.8}$ & $66.0^{+6.0}_{-6.5}$ & $76.3^{+10.9}_{-17.2}$ & 60\\
MR~2251-178  & $3.6^{+9.3}_{-9.3}$ & $9.0^{+4.0}_{-3.5}$   & $-2.1^{+4.2}_{-4.4}$ & ---               & 190\\ \hline
\end{tabular}
\end{table}

\subsection{Optical--Near-IR Flux--Flux Plots}

 \begin{figure}
 \centering
 \includegraphics[scale=0.5,angle=-90]{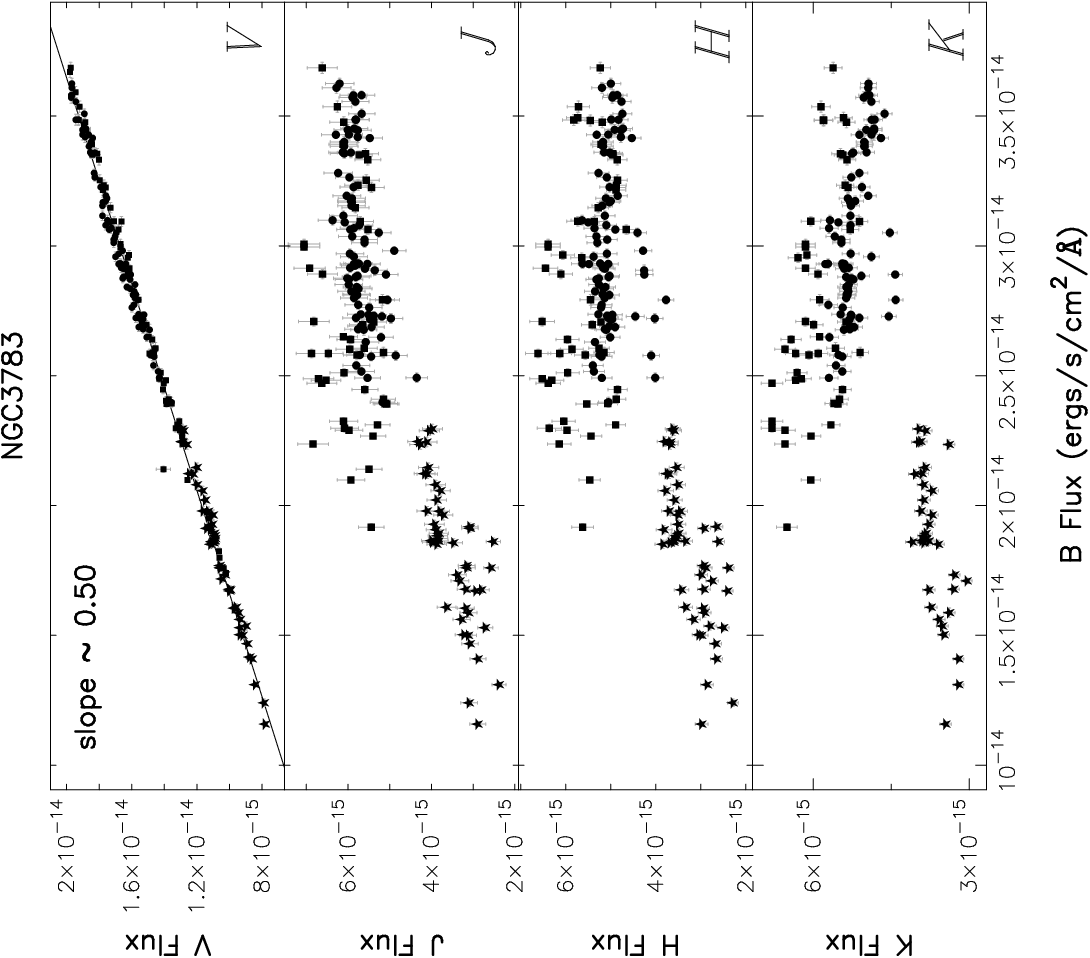}
 \includegraphics[scale=0.5,angle=-90]{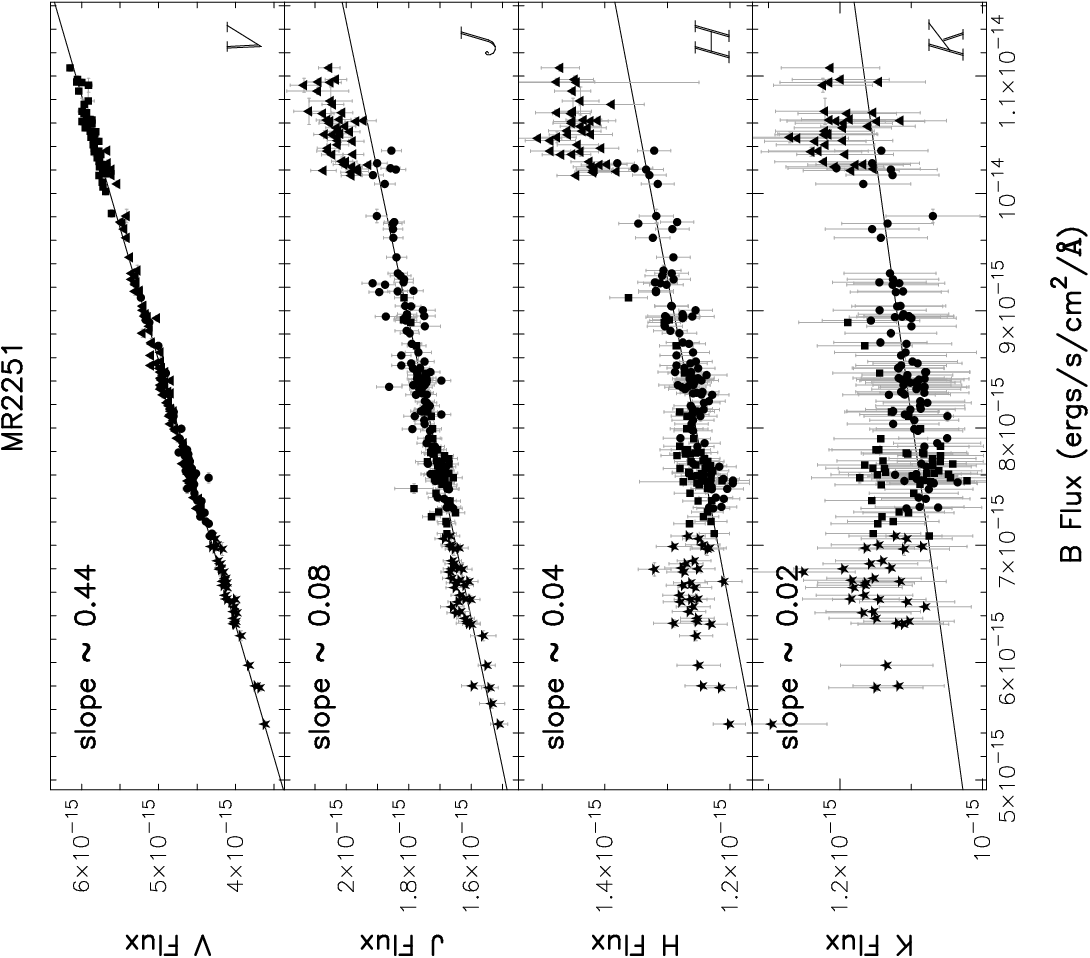}

 \caption{Flux-flux plots for NGC~3783 ({\bf top}) and MR~2251-178 ({\bf
     bottom}). Linear fits between the B and V-bands are shown for both
   galaxies. For MR~2251-178 linear fits were also determined between the B
   and the near-IR bands in the $7.5\times10^{-15}\!  <\! f_{B}\!  <\!
   10^{-14}$ ergs/s/cm$^2$/\AA\ range. Symbols correspond to the different 
   years of the monitoring campaign (see Figure 1).}
 \end{figure}

Flux-flux plots of simultaneous data points at different wavelengths can also
be used to establish the level of correlation between different emitting
regions. Figure 3 presents flux-flux plots for NGC~3783 and MR~2251-178. While
NGC~3783 has been corrected for host contribution, no correction has been
applied to MR~2251-178 given the uncertainty in this value.

To visualise a simple situation, let us assume that the emission
observed in two bands arises as simple thermal reprocessing of the
emission from a central variable X-ray source located at a height
$H_{\rm x}$ above the axis of an axisymmetric thin disc. For $R$ much
larger than $H_{\rm x}$ each radius will radiate as a Black Body of
temperature $T(R) \propto R^{-3/4}$, with most of the flux being
emitted within a narrow spectral region characterised by $\lambda
\propto T^{-1}$. Then, {\em if\/} the light travel time between these
two regions is small compared to the variability time-scales and the
sampling interval, a close-to-linear relation should emerge between
the bands.

A very tight linear relation is seen between the B and V bands for
MR~2251-178, as shown in Figure 3 (see also Ar\'evalo et al., 2008),
which is a good indication that all of the optical flux arises from a
rather compact region which is causally connected. We will assume that
this is the accretion disc.

In the previous section we determined that for MR~2251-178 the delay
between the optical and the J and H-band light curves is very
short. This is consistent with the flux-flux plots presented in Figure
3. All the near-IR bands also show linear correlations with the B-band
flux albeit with nearly flat slopes. These linear relations can be
interpreted as being due to nearly simultaneous variation in the
optical and near-IR bands. Hence, we can assume that the near-IR flux
is also being produced in the accretion disc.

We need to understand the different slopes, however. Assuming the
simple model described above, for a constant albedo we would expect
that the same {\em fraction\/} of the incident X-ray flux will be
thermalised at each radius, with the incident flux being $\propto
R^{-3}$ for large $R$. This heating flux will be added to the
gravitational energy released at each radius, which is also $\propto
R^{-3}$. Hence, we can expect that the {\em fractional\/} reprocessed
X-ray flux should remain fairly constant with radii (i.e., as a
function of wavelengths). However, our observations show that while
the B-band flux nearly doubled during the observational campaign, the
V-band presented a 50\% flux increment, and the near-IR bands show
less than 10\% variation. One possible explanation is a wavelength
dependent albedo. Another possibility is that the variation in the
near-IR is highly diluted by another near-IR component, like the
emission from a dusty torus.

It is interesting to notice that the linear relations seen between the
optical and near-IR bands for MR~2251-178 break for low and high
B-band fluxes, while they hold for $7.5\times10^{-15} \la f_{B} \la
10^{-14}$ ergs/s/cm$^2$/\AA. Outside this range, an ``excess'' of
emission appears, which can be interpreted as a new component to the
near-IR emission.  However, while the ``new'' component at low B-band
fluxes seems to be present only in H and K, the ``new'' component at
high B-band fluxes is clearly visible in all near-IR bands. This might
indicate that the near-IR excess at low B-band fluxes is due to the
presence of a dusty torus, since it is expected that the emission from
this component peaks somewhere in the mid-IR. The excess at high
B-band fluxes is consistent with the different variability trends seen
in the fourth year of monitoring, as already commented in Section
5.1. This might correspond to a new component of near-IR emission,
like an outburst in the outer parts of the accretion disc, for
example.
 
For NGC~3783 the situation is quite different due to the delays
already discussed in Section 5.2. In Figure 3 it can be seen that some
linearity is present in the flux-flux plots, but with large scatter
for all near-IR bands. In fact, two regimes are present: in the first
year of monitoring the slope of the correlations in all bands are
positive due to the consistent flux decline observed during this
period; in the second and third year of monitoring the flux-flux
slopes change from being slightly positive in the J-band to slightly
negative in the K-band. This is because of the anti-correlation in the
light curves around MJD $\sim 4520-4560$, $4600-4670$, and
$4835-4950$, where the B-band light curve presents a flux decline
while the near-IR light curves show a flux rise. Since the delay
becomes larger at longer wavelengths, the strongest anti-correlation
is seen in the K-band.

We constructed delayed flux-flux plots of the B-band versus the J and
H bands, meaning pairs of photometric data where the near-IR points
corresponded to the time of the B-band observation plus a delay. For
the J-band different delays were tried because of the very broad peak
seen in the cross-correlation plot shown in Figure 2, which could be
due to the presence of more than one emitting region, as we already
have discussed. For the H-band we tried delays within the range of
values allowed by the cross-correlation results. The delayed flux-flux
plots showed some reduction in the scatter of the photometry around a
linear fit to the two regimes observed in Figure 3, but this was
consistent with a reduction due to the smaller number of points
available when constructing the delayed photometric pairs of data, so
no further conclusions can be reached on this point.

\subsection{Modelling of the Spectral Energy Distributions}

 The nuclear BVJHK Spectral Energy Distributions (SEDs) for NGC~3783
 and MR~2251-178 are presented in Figures 4 and 5. The SED points have
 been computed as the mean of the photometric fluxes determined for
 each nucleus after resampling the intensive part of the light curves
 to the average sampling. The data have been corrected for foreground
 extinction and the host contribution is taken into account. For
 NGC~3783 this correction is fairly accurate, while for MR~2251-178 we
 have adopted a correction of 30\% of the bulge total fluxes in BJHK
 as determined in Section 4.2.

 The small number of SED points does not allow for a real fitting
 procedure; instead we have attempted to use all available information
 for each source and modelled the SED taking into account these
 constraints. These are: 1) the SED itself; 2) the knowledge that the
 disc and/or another emitting region contributes significantly to the
 individual SED points; 3) the restrictions on the location of the
 different emitting regions from the cross correlation analysis
 presented here and in Ar\'evalo et al.\ (2008, 2009); 4) the spectra
 of the fast variability component, which we assume is modulated by
 the reprocessing of the X-ray emission on the disc and the intrinsic
 variations in the accretion flow; 5) the spectra of all the
 variability signal, where any constant component to the SED has been
 removed. 

 We want to build a toy model that can account for the observational
 constraints imposed by our sources. We assume a model of an
 $\alpha$-disc illuminated by a central X-ray source. Analytically,
 the emitted SED corresponds to the sum of the Black Body emission
 ($B_{\nu}$) from each of the disc annuli:

\[ F_{\nu} = \int^{R_{out}}_{R_{in}} B_{\nu}(T(R))\ d\Omega \]

where $R_{in}$ and $R_{out}$ are the innermost and outermost disc
radii and $d\Omega$ is the solid angle subtended by a disc annuli as
seen by the observer. The temperature as a function of radius $T(R)$
is found to be (Cackett, Horne \& Winkler, 2007):

\begin{small}

\[ T(R) = \Big\{  \Big( \frac{3GM \dot{M}(1\!-\!\sqrt{R_{in}/R})}{8 \pi R^{3} \sigma \eta \sqrt{1\!+\!H'(R)^{2}}} \Big) + \Big( \frac{(1\!-\!A)L_{\rm x}}{4 \pi R_{\rm x}(R)^{2}} \Big) \cos \theta_{\rm x}(R) \Big\}^{1/4} \]

\end{small}

\begin{table}
\caption{Parameters for Spectral Energy Distribution Models. Masses
  are given in units of $M_{\odot}$ and accretion rates in Eddington
  units, with $\eta$ corresponding to an accretion efficiency of
  0.1. The disc profile is described by a power law of the form
  $(R/R_a)^{\beta}$, with a flare radii $R_{a}$ (in units of
  $R_{g}=GM/c^2$). The height of the X-ray source ($H_{\rm x}$) is
  given in units of $R_{g}$, and the Torus Black Body temperatures
  ($T$) in Kelvin. $\dag$ For Model B in NGC~3783: $\beta$ and $R_a$
  correspond to the characterisation of the disc 'hump'. $\ddag$ For
  Model B in MR~2251-178: $R_t$ represents the radius of transition
  between the flared and tapered disc profile (in units of
  $R_{g}$). All the listed parameters, with exception of the black
  hole masses, are considered as free variables for the modelling.}
\begin{tabular}{lccccccc} \hline
Galaxy & $M$ & $\dot{M}/\eta$ & $\beta$ & $R_a$ &$(1\!-\!A)L_{\rm x}$ & $H_{\rm x}$ & $T$ \\
Model & & & & $R_t$ & (ergs/s/cm$^2$) & & \\  \hline
\multicolumn{2}{l}{NGC~3783}& & & & & & \\ 
Model A   & 3e$10^7$ & 0.015 & 1.0& 10 & 3.8e$10^{43}$ & 50 & 1200\\ 
Model B   & 3e$10^7$ & 0.015 & 1.0& 10 & 3.8e$10^{43}$ & 50 & 1200\\ 
          &          &       & 1.3$^\dag$& 50$^\dag$ & & & \\ 
\multicolumn{2}{l}{MR~2251-178}& & & & & & \\ 
Model A & 3e$10^8$ & 0.01 & 1.8& 100 & 5.9e$10^{44}$ & 50 & --- \\
Model B & 3e$10^8$ & 0.01 & 1.0& 1   & 4.5e$10^{44}$ & 80 & 1500\\ 
          &          &    &    & 100$^\ddag$ & & & \\ 
\hline
\end{tabular}
\end{table}

 \begin{figure*}
 \centering
 \includegraphics[scale=0.65,angle=-90]{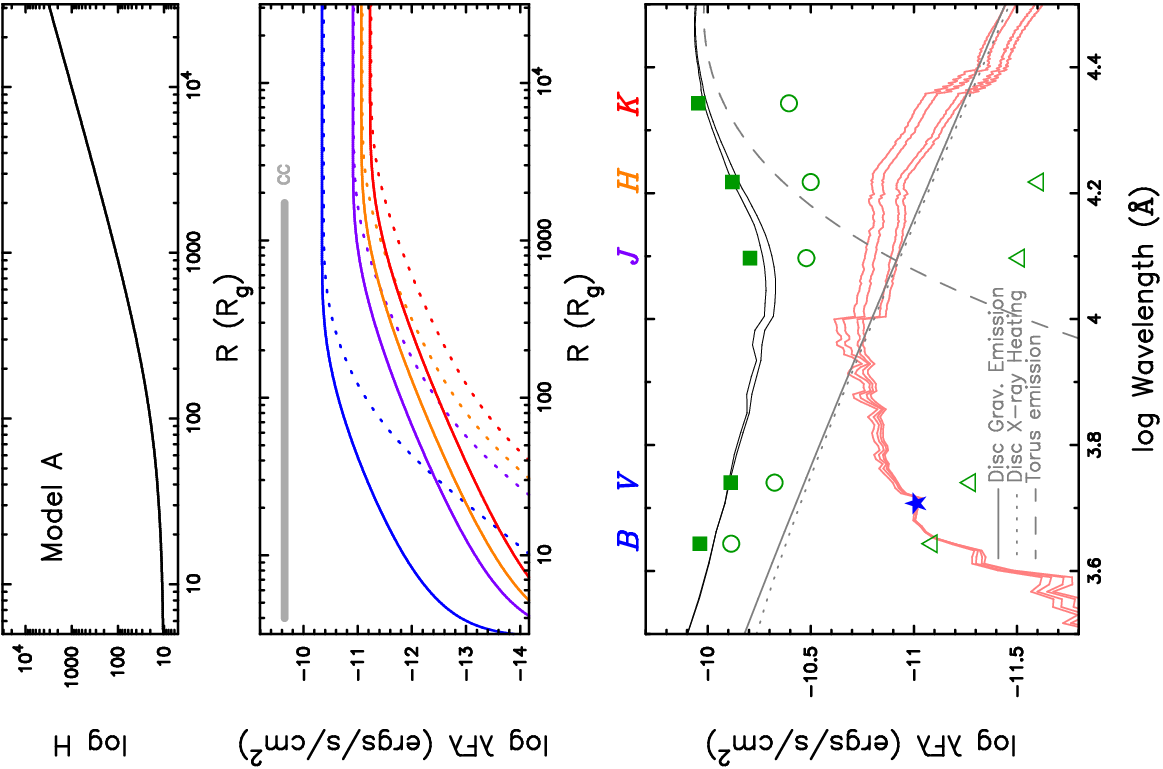}%
 \hspace{0.4cm}
 \includegraphics[scale=0.65,angle=-90]{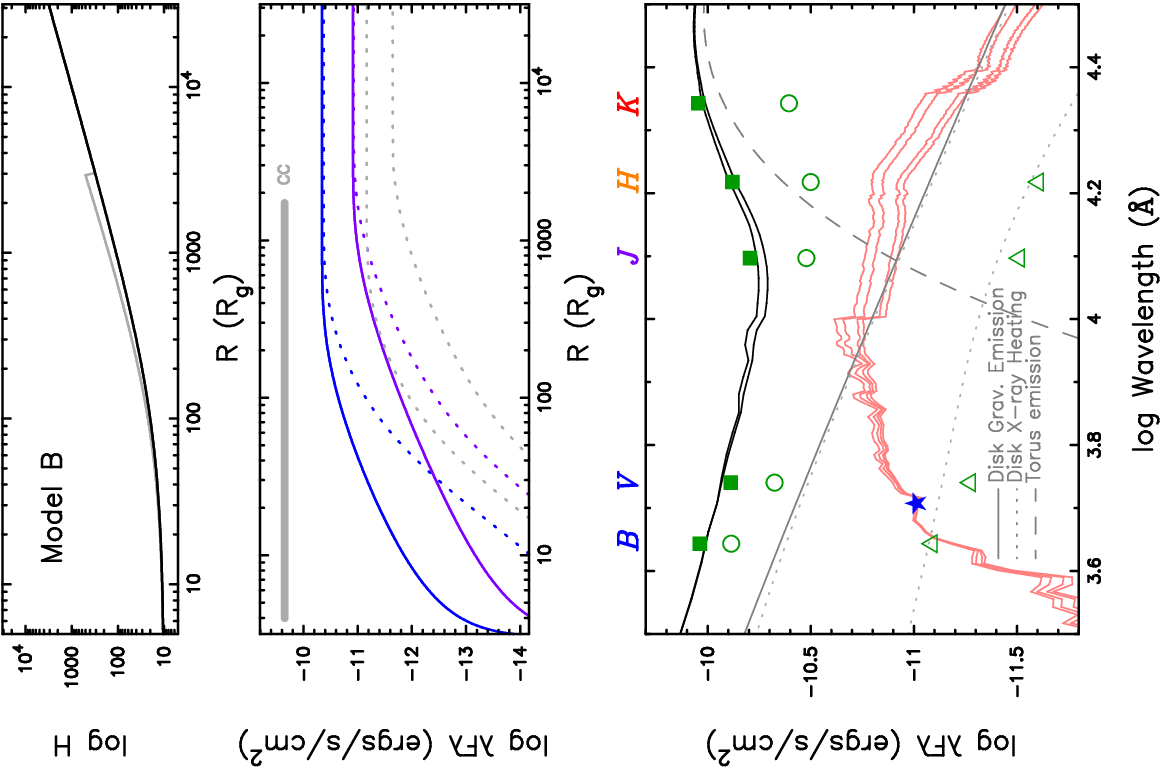}%
 \caption{$\alpha$-disc models for NGC~3783: Model parameters are
   presented in Table 4. {\bf Top} panels show the disc profile $H$ as
   a function of radius $R$ in units of $R_g$. {\bf Middle} panels
   show the cumulative emission in the BJHK-bands as a function of
   radius. The solid lines represent the emission due to the release
   of gravitational energy, while the dotted lines represent the
   emission due to the reprocessing of X-rays (in the {\bf right
   middle} panel only the B and J-bands are presented, while the
   reprocessed emission of the X-rays by a `hump' is included in
   light-grey colour with its profile included in the {\bf right
   upper} panel). The light-grey ``CC'' horizontal bar represents the
   distance derived from the cross correlation between the X-ray and
   the B-band emission (Ar\'evalo et al.\ 2009). {\bf Bottom} panels
   show the optical and near-IR observed SEDs ({\bf solid
   squares}). {\bf Empty circles} represent the difference SED between
   the maximum and minimum flux measurements, while {\bf empty
   triangles} represent the flux rms of the fast variability. The {\bf
   continuous black lines} represent the total sum of the different
   model components, which are: the gravitational and X-ray
   reprocessed disc emission (in the {\bf right bottom} panel emission
   from the 'hump' is shown separately), a Black Body representing the
   Torus contribution, and the host contribution, scaled to the
   measurement at 5100 \AA\ ({\bf star}). The host has been
   represented by four single stellar populations with $Z=1, 2\
   Z_{\odot}$ and 9, 12 Gyr of age. A jump at $\log(\lambda)=4$ has
   been introduced to account for the change in the aperture used for
   the photometry in the optical and near-IR data. The total fit was
   determined using the two most extreme stellar populations which are
   characterised by $Z=1\ Z_{\odot}$, age=12 Gyr and $Z=2\ Z_{\odot}$,
   age=9 Gyr.}
 \end{figure*}

 \begin{figure*}
 \centering
 \includegraphics[scale=0.65,angle=-90]{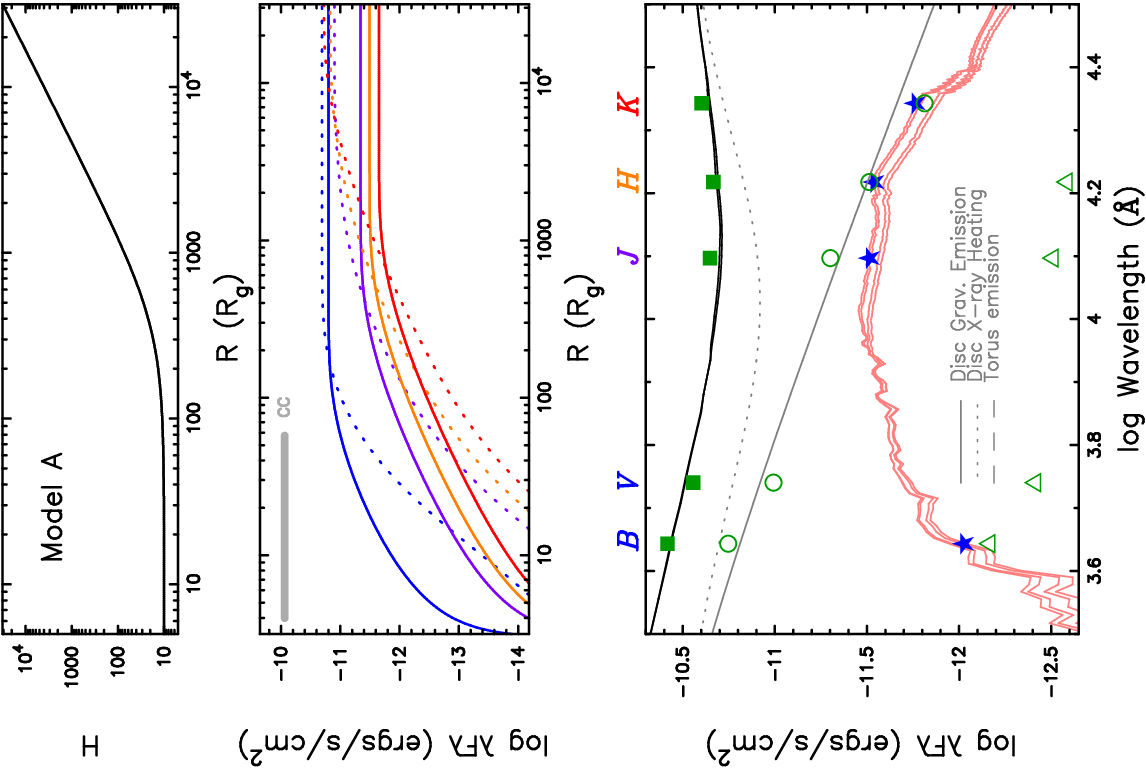}%
 \hspace{0.4cm}
 \includegraphics[scale=0.65,angle=-90]{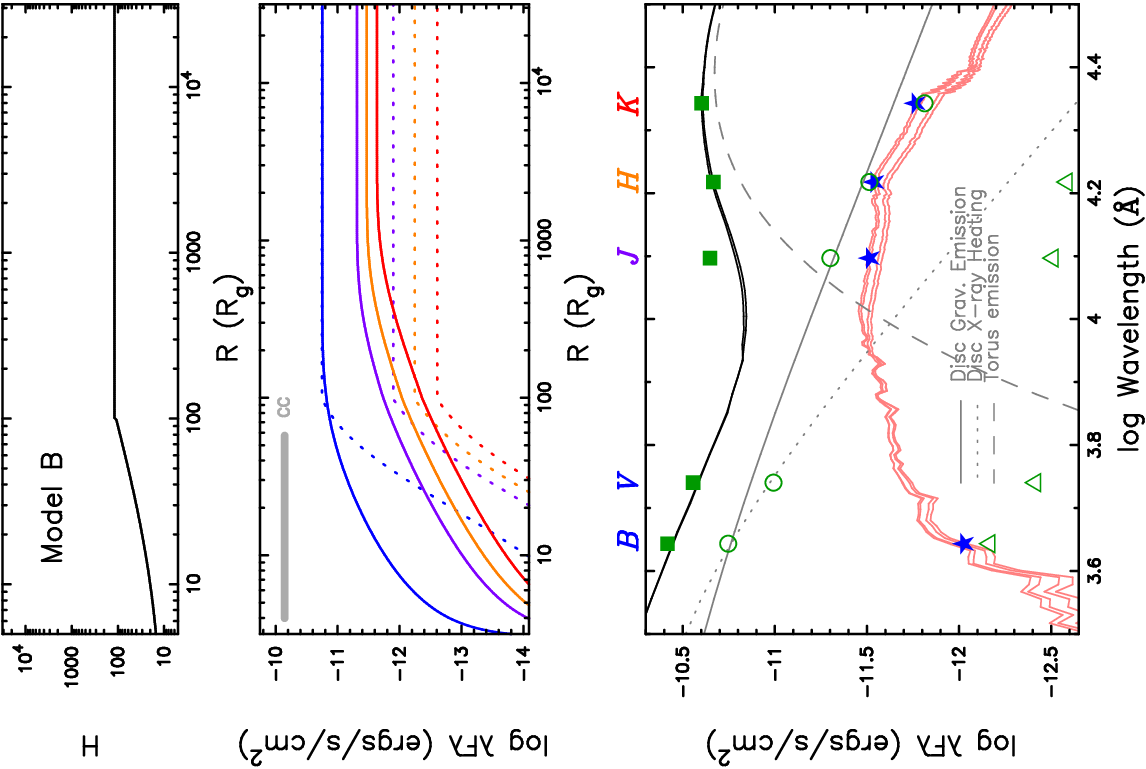}
 \caption{$\alpha$-disc models for MR~2251-178. Model parameters are
   presented in Table 4. Panels and symbols are the same as in Figure
   4. The {\bf left} and {\bf right} model SEDs present models without
   and with a Torus component (modelled as Black Body emission),
   respectively. The host contribution has been scaled to the
   estimated bulge fluxes in BJHK as determined in Section 4.2
   assuming a 30\% contribution of the bulge light into the used
   photometry apertures (stars).}
 \end{figure*}

where $\dot{M}$ is the accretion rate, $\sigma$ is the Stefan-Boltzman
constant, $\eta$ the accretion efficiency, $H'(R)$ is the derivative of the
function that describes the vertical profile of the disc, $A$ is the disc
albedo, $L_{\rm x}$ is the 0.01--500 keV luminosity of the central X-ray
source, $R_{\rm x}$ is the distance from the X-ray source to a position at
radius $R$ on the disc, and $\theta_{\rm x}$ is the angle between the
direction on the incident X-ray radiation and the local normal to the disc
surface. Since the true factor to extrapolate from the observed X-ray
luminosity to the 0.01--500 keV energy range is not well known, nor is the
disc albedo, the $(1\!-\!A)L_{\rm x}$ term can be regarded as a ``fudge'' term
in the above equation. Other model parameters are the height of the X-ray
source above the disc ($H_{\rm x}$) which determines $\theta_{\rm x}$ and
$R_{\rm x}$, and the disc central height which we always assume to be equal to
10 $R_g$ (notice that this introduces an asymptotic behaviour in the
logarithmic representation of the disc profile for small values of $R$, as can
be seen in the top panels of Figures 4 and 5). Model parameters for our
sources are presented in Table 4, all of which are allowed to vary freely
except for the black hole mass. We did not perform a residual minimization
routine to fit the model, given the small number of data points and large
number of free parameters; we simply searched for parameters that produced a
good description of the data. Therefore, no error estimation of the parameters
is possible.

The disc emission is represented in the SEDs by two components
(Figures 4 and 5, bottom panels): one due to the release of
gravitational energy only (first term in the previous equation), and
another representing the emission due to the reprocessing of X-rays
(second term in the previous equation). To the disc emission we have
added the near-IR emission from a warm dusty torus. No direct evidence
for the presence of a torus is found in our data for MR~2251-178.
However, the SED fit suggests that torus emission might be present in
this object, as discussed further in Section 5.4.2. Figures 4 and 5
also present the cumulative emission for different bands from the
release of gravitational energy and the reprocessing of X-rays (middle
panels) and the disc profile as a function of radius (top panels).

Ar\'evalo et al.\ (2008, 2009) have shown that the optical long term
variability (on time scales of years) is most likely driven by
fluctuations in the accretion flow. On the other hand the component
corresponding to the reprocessing of X-ray heating carries the fast
modulation from the variable X-ray central source. Hence, we can
estimate a lower limit to the flux from this component by measuring
the rms of the fast variability. To isolate the very fast variations
we have computed the flux rms of the residuals of the yearly segments
of the light curves with respect to a fit using a low order (up to
quadratic) Legendre polynomial in the V, B, J and H bands. These flux
upper limits are plotted in Figures 4 and 5 using triangles. The
K-band was omitted because of the larger errors.

We also determined the difference spectrum for the sources, this is, the
spectra obtained by subtracting the observed SED around the lowest flux level
from the spectra observed around the highest flux level. When computing the
difference spectra any constant component to the SED will be subtracted
out. In particular, the host galaxy contribution should be completely
eliminated for both sources, and since no clear evidence for the presence of
variable emission from the torus in MR\,2251-178 is found from the
cross-correlation analysis, its difference spectra should represent only disc
emission. In general, any disc component that remained constant during the
period of monitoring will also be subtracted out, while any component with
constant shape but different scaling factors will retain its spectral
signature.

A final remark: since there is now good evidence that some (and more
likely most) of the variability is driven by changes in $M_{\odot}$,
it is possible that the gravitational heating component could vary by
a factor of a few with $\log R$, while our toy models, which assume a
steady state approximation, are not capable of reflecting this dynamic
behaviour.

\subsubsection{NGC~3783}

The near-IR constraints we would like to meet for NGC~3783 are a
non-negligible amount of disc emission in the J-band and a torus component
that becomes progressively more significant when moving from the J to the
K-band. Variability constraints also tell us that the optical emission
presents a 6 day lag behind the X-rays (Ar\'evalo et al., 2009). The
difference spectrum presented in Figure 4 (shown with empty circles in the
bottom panel) corresponds to emission from all the variable components in
NGC~3783, and presents the expected upturn in the near-IR, as the variable torus
emission becomes prominent in the H and K-bands. On the other hand, the very
fast variability (empty triangles), which as argued before should be dominated
by the X-ray reprocessed emission, does not present such turn up.

Following Ar\'evalo et al.\ (2009), we first model the disc in
NGC~3783 as a surface with constant radius-to-height ratio, $R/H = 10$
(using a power law notation, $H(R) = (R/R_a)^{\beta}$, with $\beta=1$
and $R_a=10$). As can be seen in the bottom panels of Figure 4,
the components corresponding to the release of gravitational energy
and X-ray heating have very similar spectral slopes all the way from
the B to the K-band and therefore their scaling is not constrained by
the modelling. Adopting a 2--10 keV X-ray luminosity for NGC~3783 of
$1.3\times10^{43}$ ergs/s, a conversion factor to the 0.01--500 keV
band of 5, and an albedo of 40\%\ as determined by Ar\'evalo et al.\
(2009), we obtain the values for $\dot{M}$ and $(1\!-\!A)L_{\rm x}$
presented in Table 4. Allowing for the presence of a dusty torus with
a temperature of 1200 K, it is possible to obtain a good fit to the
observed SED, as seen in Figure 4. The J-band presents similar
contributions from the disc and torus, while the torus becomes more
prominent towards longer wavelengths.

The unexpectedly long 6 day lag between the X-rays and the optical emission is
shown in the middle panels of the Model A in Figure 4 as a long light-grey
``CC'' horizontal bar, which extends from the axis of symmetry of the system,
where the X-ray source is assumed to be located, to a radius of several light
days ($R \ga 1000\ R_g$).  Even though the formal errors for the lag between
the X-ray and optical bands allow for a 2 sigma distance $R \la 100\ R_g$,
Ar\'evalo et al.\ (2009) argue that this is a robust result since the positive
lag was found when obtaining the cross-correlation of the entire light curve
as well as when using the period of fast monitoring.

Model A is not consistent with the 6 day lag between the X-ray and
optical emission. We do not know with certainty what fraction of the
optical flux is causing the variability which gives rise to the cross
correlation signal, and although the fast variability measurements
(green triangles in Figure 4) suggest that this might not be very
large, Model A shows that hardly any optical emission is produced past
$R \sim 500\ R_g$, which contradicts our previous results.

The 6 day lag is driven by the reprocessing of X-ray emission, since the
optical bands vary behind the X-rays, as this emission illuminates and heats
the accretion disc. To shift the amount of X-ray reprocessed emission to
regions further out from the central source we need to introduce some flared
structure that would increase the surface area that receives direct
illumination from the central X-ray source. However, to have a significant
amount of optical emission at $R \ga 1000\ R_g$ the X-ray heating has to be
very efficient so that high enough temperatures are obtained. If this is
achieved assuming an azimuthaly symmetric, $2\pi$, flaring geometry the total
amount of reprocessed emission, from the optical to the near-IR, exceeds the
observed SED by several orders of magnitude.

Therefore we assume that superimposed onto the geometry already
presented in Model A a `hump' (a spiral arm?) located at $R \sim 1000\
R_g$ and subtending a very small solid angle is heated to very high
temperatures and produces the optical emission that lags the X-ray
central source by 6 days. The smaller subtended area of this structure
also alleviates the problem of observing a coherent variation from a
structure located at such large distances from the central
region. This scenario is presented by Model B in Figure 4, which
is a small departure from model A and therefore retains several of the
previous traits, as can be seen in Table 4. As before, emission from
a dusty torus is also included. Light-grey curves show the
characteristics of the new hump component. The general fit to the SED
is still good, as this new component is not energetically very
significant.

The presence of the hump is visible in the top panel as two radial
cuts are shown, one running across it and the other one running across
any other radial direction. To avoid overcrowding, the middle panel
only shows the emission from the general disc in the B and J-bands,
while the emission from the reprocessing of the X-rays by the hump,
also in the B and J-bands, is included in light-grey colour. The hump
shifts some of the reprocessing of the X-ray emission towards
distances more consistent with the cross correlation results and its
SED is consistent with that of the fast variability. Most likely, to
properly model this component we would have to introduce an
non-axisymmetric geometry, which is beyond the scope of our simple toy
models.

\subsubsection{MR~2251-178}

No direct evidence for the presence of a torus is found in the
lightcurves of MR2251-178 (see also Section 6.2 for further analysis),
while disc emission is clearly seen from the optical to the near-IR
bands, as determined by our analysis of the cross-correlation and
flux-flux plots (Ar\'evalo et al., 2008; Section 5.2). In fact, the
cross-correlation results set constraints to the distance between the
regions of the disc responsible for the variable emission at optical
and near-IR wavelengths, with no more than a few days as upper
limits. In Figure 5 the distance from the centre of symmetry of the
system to a radius of 1 light-day is represented by a ``CC''
horizontal bar. Also, for MR~2251-178 the difference spectrum
presented in Figure 5 is steep all the way from the optical to the
near-IR, while the fast variability spectrum, which is dominated by
the X-ray reprocessing, shows a slightly flatter spectral shape.

Table 2 shows that the variable component in the near-IR light curves
of MR~2251-178, either driven by the reprocessing of X-ray emission or
the intrinsic fluctuations in the accretion flow is small. This is
also shown by the very flat slopes in the flux-flux plots presented in
Section 5.3. Therefore, the near-IR emission is dominated by a
non-variable component and the variability constraints have to be used
with care, since they might not represent what the bulk of the near-IR
emission is doing. More valuable information is found in the
difference and the fast variability spectra, as we will see next.

Without direct evidence for the presence of a dusty torus from our cross
correlation analysis, we want to consider how to produce the large amount of
near-IR flux shown in the SED using the accretion disc alone. To produce
near-IR emission from the disc we adopted a disc profile of the form
$(R/R_a)^{\beta}$ with $\beta = 1.8$ and $R_a = 100 R_g$ to achieve the
desired flaring geometry. The result can be seen in Model A of Figure 5. The
general fit to the observed SED is reasonably good and the emission at optical
wavelengths peak close to the limit set by the ``CC'' bar. However, the X-ray
reprocessed component presents a spectral shape which is very different to
both, the difference spectrum and the fast variability spectrum, while the
near-IR bands peak at $R > 1000 R_g$. In fact, the model predicts that $\la
90\%$ of the K-band emission corresponds to reprocessed X-rays and this large
output should introduce larger variability than that observed. For example,
$\sim 50\%$ of the reprocessing is taking place within $\sim 3000 R_g$, which
corresponds to a light travel time of $\sim 30$ days. Therefore the K-band
light curve should be very similar to that of the optical emission but
smoothed with a kernel of 30-60 days (60 days corresponding to an edge-on
line-of-sight towards the disc), which is not the case. Finally, to achieve
the necessary geometry, an extremely thick disc is required, with $H \ga R$
for $R \ga 10^4 R_g$, which is not expected for the estimated accretion rate
observed in this object (see Section 6.3). Hence, we discard this model.

The non-variable near-IR emission could originate in a dusty torus
instead of the disc itself, which is bound to be more variable.
Hence, we will explore a model where the near-IR emission arises from
a non-variable dusty torus. 

First, we assume that a dusty torus is responsible for most of
the near-IR emission, while most of the variability signatures should
originate in the disc. However, no simple flaring geometry would match
both, the shape of the SED and the time constraint from the
cross-correlation analysis, which imply a surprisingly fast near-IR
response. Hence, in Model B we introduce a further modification to the
model geometry, with the disc presenting a tapered profile after a
characteristic radius $R_t$. Effectively, the tapering (in this case
to a slab profile) stops the X-ray heating for $R > R_t$. This ensures
that the bulk of the near-IR reprocessed emission comes from regions
inwards of $R_t$. Also, the tapering stops the rapid increase in the
disc thickness introduced by the strong flare required at low $R$.

Our fit with a flaring $+$ tapered geometry characterised by $\beta =
1.0, R_a = 1 R_g, R_t = 100 R_g$, and emission from a torus is
presented in the right panel in Figure 5. Other parameters are found
in Table 4. The fudge luminosity for Model B corresponds to $4.5
\times 10^{44}$ ergs/s. Orr et al.\ (2001) estimated a 0.01--500 keV
X-ray flux for MR~2251-178 of $1.75\times10^{-10}$ ergs/s/cm$^2$,
while Ar\'evalo et al.\ (2008) inferred an albedo of 0.4 for the disc
surface. For a distance of 274 Mpc, this results in a total X-ray
luminosity of $9.5 \times 10^{44}$ ergs/s. Our model therefore, is a
factor of 2 too low when compared with this prediction. However, the
SED shows that there is still some degree of degeneracy between the
scaling of the $\dot{M}$ and $(1\!-\!A)L_{\rm x}$ components, while
the actual albedo might be larger than predicted. More problematic is
the fact that the spectrum of the X-ray reprocessed emission is very
steep, as a result of the truncation of further near-IR reprocessing
at larger $R$. This is in sharp contrast with the flat spectrum of the
fast variability component, particularly in the near-IR (although the
fast variability in the H-band is poorly constrained for this object -
see Figure 1). No combination of the (already quite large) set of
model parameters could alleviate this problem.

\section{Discussion}

\subsection{MR~2251-178: disc or jet emission?}

One possible origin for the nearly simultaneous optical and near-IR
variability in MR~2251-178 is the presence of a jet. In this case the
emission would be due to synchrotron instead of thermal processes. Jet
emission is found to be prominent in radio-loud AGN and in
relativistically boosted radio-quiet AGN (also known as
radio-intermediate AGN; Falcke et al., 1996; Barvainis et al.,
2005). Radio-loud AGN are defined as those with $R_L > 100$ (where
$R_L=f_{5 GHz}/f_{B}$; Kellermann et al., 1989), radio-intermediate
AGN are those with $3-10 \la R_L \la 100$, while radio-quiet AGN have
$R_L <<10$.

Radio counterparts to MR~2251-178 have been found from the
cross-correlation of the NRAO VLA Sky Survey (NVSS) with the Second
Incremental Data Release of the 6 degree Field Galaxy Survey (6dFGS
DR2; Mauch \& Sadler, 2007) and the ROSAT Bright Source Catalog (RBSC;
Bauer et al., 2000).  The determined flux is about 16 mJy at 1.4
GHz. Assuming a radio spectral index $\alpha = -0.2$ for
radio-intermediate AGN ($S_{\nu} \propto \nu^{\alpha}$; Barvainis et
al., 2005) we can infer the flux at 5 GHz. Hence, we find that
MR~2251-178 has $R_L \sim 0.14$ based on our observed B-band $<\!f\!>$
flux (Table 2). Clearly, MR~2251-178 corresponds to an unbeamead
AGN. Notice that the $\alpha$ value for radio-quite sources can be
even steeper, making the derived $R_L$ even smaller.

\subsection{A torus in MR~2251-178?}

\begin{figure}
 \centering
 \includegraphics[scale=0.3,angle=-90]{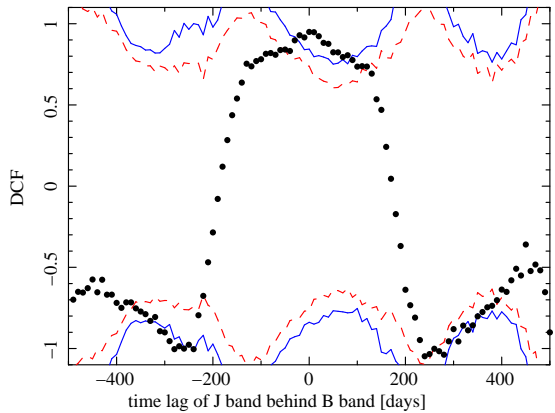}%
 \caption{Discrete Correlation Functions for MR~2251-178 between the B
 vs J bands for a 500 day span. Dash and continuous lines are the same
 as in Figure 2.}
\end{figure}

\begin{figure}
 \centering
 \includegraphics[scale=0.55,angle=0]{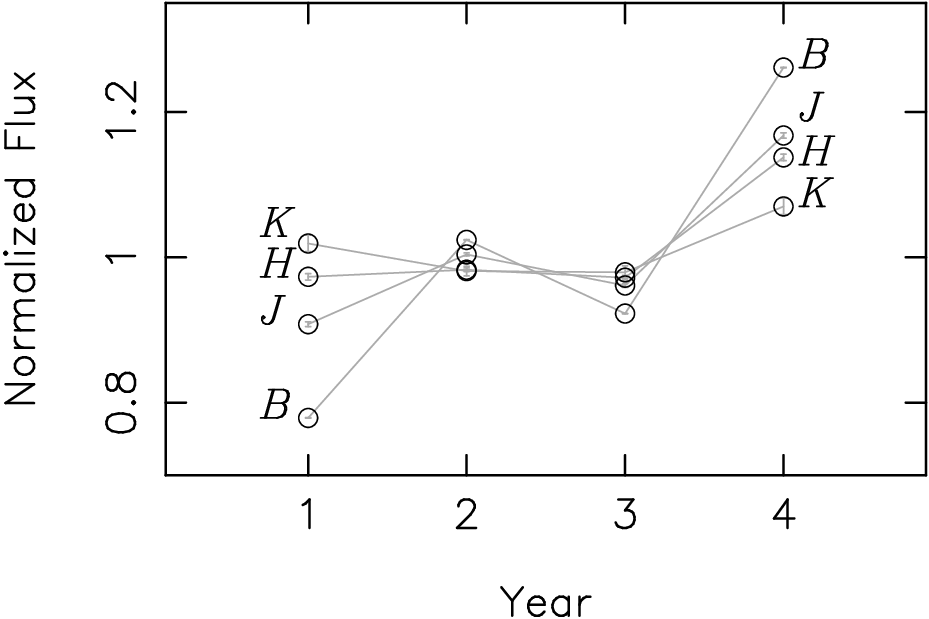}%
 \caption{Yearly averaged and normalised B, J, H and K-band
 light-curves for MR~2251-178. A 30\% flux contribition from the bulge
 component has been assumed (see Section 4.2). }
\end{figure}

No evidence for a torus is seen in the cross correlation analysis
presented in Section 5.2, which indicates that the response from all
near-IR bands to the variable X-ray source seems to come from a much
closer source, that we identify with the accretion disc. However,
since the Suganuma et al.\ (2006) relation predicts the presence of a
torus at the dust sublimation radius which, for MR~2251-178 is at 190
light days ($R \sim 1000$ in units of $R_{g}$), we need to explore the
DCF to cover longer time scales.

Figure 6 presents the long-term cross-correlation between the
optical and J band. In this case we did not calculate separate DCF
functions for each yearly segment and average the results but rather
calculated the DCF of the entire light curves. For this reason, the
large amplitude long-term fluctuations dominate the correlation,
making the central peak much broader. The good correspondence between
smaller-amplitude, rapid fluctuations, which dominated the DCF in
Fig. 2, produces the narrower peak superimposed on the center of the
distribution, again consistent with a zero lag. No positive
correlation features are found at the expected dust sublimation lag of
190 days. The result obtained using the H band is of worse quality but
consistent to that presented in Figure 6, while using the K band gives
unconstrained results, as expected.

The low signal-to-noise of the K-band light curve is in fact
preventing us from better assessing the presence of the torus at the
wavelength where it should make a more significant
contribution. Figure 7 presents the normalised yearly averaged light
curves for MR~2251-178 which allows us to examine the long term
variations in each band, minimizing the impact of photometric
errors. It can be seen that the variability pattern changes towards
longer wavelengths, where the variations are not only smaller in
amplitude, but also move in the opposite sense with respect to the
shorter wavelengths: between the first and second year of monitoring
the flux increases at short wavelengths but decreases in the K
band. This might be the signal of a slowly varying torus.

The lack of clear variability from the torus, however, is
surprising. In type 1 objects we should have a direct view to the
central source and the inner face of the torus, which is the region
where the largest amount of variation should be found. Maybe the torus
is located further away than expected and the variation should be seen
at even longer time scales than the ones we are able to study. This
might be the case if MR~2251-178 was significantly brighter in the
past, forcing the sublimation radius to recede to large distances.
Similarly, a variation in the distance to the inner part of the torus
in NGC~4151 was reported by Koshida et al.\ (2009).

Alternatively, the torus geometry might subtend a small solid angle as
seen by the central source, in which case its varying flux might be
buried in the disc emission.  In fact, a near-IR ``excess'' is seen in
the flux-flux plots presented in Figure 3, but only in the H and
K-bands. This corresponds to observations during the first year of
monitoring. Unfortunately only a monotonic decrease in flux is
observed at all wavelengths during that period, so if this emission
comes from the torus it is not possible to draw any conclusions about
its possible response to variations from the central source.

From the analysis of XMM-Newton data Kaspi et al.\ (2004) found that
the 6.4 keV iron emission line in MR~2251-178 is weak (with an
equivalent width of $53 \pm 20$ eV), suggesting a weak torus
component. However, this is also consistent with a high albedo if the
accretion disc is very highly ionised which is a result also
consistent with our modelling of the disc in Section 5.4.2.

The need for Black Body emission in the near-IR is also suggested by
our fit to the SED. This is consistent with the presence of a torus,
which would dominate the SED beyond wavelengths of a few microns.

\subsection{Hot and cold disc emission}

\begin{table} 
\caption{Black Hole masses and accretion rates. Masses are given in
  units of $10^{6}M_{\odot}$ and $\dot{M}$ in Eddington units. Refs.:
  1 Peterson et al., 2004; 2 Woo \& Urry, 2002, 3 this work; 4
  Peterson et al., 2005; 5 Wang, Mao \& Wei, 2009; 6 Ogle et al.,
  2004; 7 Kishimoto, private communication.}
\begin{tabular}{ccccc} \hline
Galaxy & $M$ & $\dot{M}$ & $(\dot{M}/M)^{(1/4)}$ & Refs.\\ \hline
NGC~3783      & 30       & 0.06       & 0.21 & 1, 2 \\ \smallskip
MR~2251-178   & 300      & 0.04-0.4   & 0.11-0.19 & 2, 3 \\ \hline 

NGC~3227      & 42       & 0.014      & 0.14 & 1, 2\\ \smallskip 
NGC~4395      & 0.4      & 0.0012     & 0.23 & 4 \\
NGC~4051      & 2        & 0.15       & 0.52 & 1, 5, 6 \\
NGC~7469      & 12       & $\ga 1$    & 0.54 & 1, 2\\ \hline

Q0144-3939    & 1,180    &  0.26      & 0.12 & 7\\
3C95          & 8,330    &  0.37      & 0.08 & 7\\
CTS A09.36    & 1,180    &  0.29      & 0.13 & 7\\
4C09.72       & 10,290   &  0.22      & 0.07 & 7\\
PKS2310-322   & 1,420    &  0.37      & 0.13 & 7\\
Ton202        & 4,115    &  0.15      & 0.08 & 7\\ \hline
\end{tabular}
\end{table}

Our analysis clearly indicates the detection of variable near-IR
emission from the disc of MR~2251-178 and suggests that some disc
near-IR emission is also present in NGC~3783. Evidence is groing
that accretion disc emission is indeed visible in the near-IR regime
(Landt et al., 2011; Kishimoto et al., 2008; Minezaki et al.,
2006). In fact, based on the cross-correlation analysis of the optical
and J and H-band data for NGC~4395, Minezaki et al.\ (2006) argue that
the observed variability comes from the outer region of the accretion
disk. In addition, Tomita et al.\ (2006) have assumed the presence of
a variable disc component that extends from the optical to the near-IR
to interpret the light curves of MCG~+08-11-011. They find a SED index
for this component of $\alpha = -0.1 \pm 0.4$, still consistent with
the predicted value of 1/3.

Suganuma et al.\ (2006) did not find direct evidence in any of their
sources for the presence of near-IR emission arising from the
accretion disc. In fact, their optical versus near-IR flux-flux plots
for NGC~5548, NGC~4051 and NGC~3227 (their Figure 13) look like pure
scatter diagrams, except for NGC~7469 where some correlation seems to
be present. This seems to indicate that in these sources the
variable near-IR emission is dominated by the distant large torus, or
by the much slower variations of the accretion flow in the outer parts
of the accretion disc. On the other hand, in MR~2251-178 the variable
near-IR emission is dominated by the reprocessing of X-rays by the
disc.

Uttley et al.\ (2003) already pointed out that the nature of the disc
reprocessed emission will depend on the location of those regions of the disc
exposed to the X-ray illumination. The range of temperatures that characterise
the reprocessed emission will depend on the thermal structure of the disc,
which in turn is determined by the black hole mass and the accretion rate as
$T \propto M^{-1/4}$ and $T \propto \dot{M}^{1/4}$, for $\dot{M}$ expressed in
Eddington units. Furthermore, Uttley et al.\ (2003) invoke results from X-ray
binaries to argue that the size of the X-ray emitting region, expressed in
units of $R_g$, is constant for all sources. Therefore, for small accretion
rates and/or large black hole masses, we expect a cold accretion disc, where
the near-IR emitting region will be located at small values of $R/R_g$ and
will be more likely to be exposed to a significant fraction of the X-ray
flux. The opposite situation, a hot accretion disc, is expected for sources
characterised by small Black Hole masses and/or large accretion rates. The
{\em fast\/} variability observed in different bands, which traces the
variations produced by the central X-ray source, therefore, represents a tool
that allow us to map the regions being illuminated and heated by X-ray flux.

To test this scenario we have gathered the values of the black hole
masses and accretion rates for our sources and those found in Suganuma
et al.\ (2006), Minezaki et al.\ (2006), and Kishimoto et al.,\
(2008). These are presented in Table 5. For NGC~3783 and MR~2251-178
the accretion rates are based on the observed bolometric luminosities,
and not on the values determined during our SED fitting. This is
because our determination of $\dot{M}$ only accounts for the direct
release of gravitational power into the disc, while the X-ray heating
is determined by the observed value of $L_{\rm x}$. However, the high
energy emission (and any released mechanical energy) is also powered
by the gravitational potential energy which ultimately takes the form
of a hot corona (and a mechanical jet). Since we do not have a
prescription to relate $\dot{M}$ with $L_{\rm x}$, $L_{\rm x}$ remains
as an extra parameter in the SED models.

As before, measurements of $M$ and $\dot{M}$ for NGC~3783 are fairly
accurate, while clear uncertainties remain for MR~2251-178. Besides,
the black hole estimates presented in Section 2, the determination of
the bolometric luminosity for MR~2251-178 is rather vague: assuming
the bolometric corrections computed by Marconi et al.\ (2004) we find
that the B-band luminosity predicts $L_{bol} \sim 2\times10^{45}$
ergs/s, while the 2-10 keV luminosity gives $L_{bol} \sim
2\times10^{46}$ ergs/s. This shows that MR~2251-178 is fairly ``X-ray
loud'', with an $\alpha_{ox}$ index of 1.2 (for a definition of
$\alpha_{ox}$ see Tananbaum et al., 1979), close to the edge of the
distribution for quasars and local Seyferts (Zamorani et al., 1981;
Grupe et al., 2010). Hence, there is about one order of magnitude
uncertainty in the bolometric luminosity of MR~2251-178, as well as in
its accretion rate.

Table 5 shows that MR~2251-178 is located in the lower part of the
distribution of $(\dot{M}/M)^{(1/4)}$ values when compared with local
NGC Seyferts and, if its lower accretion rate is confirmed, it fits
well within the range of $(\dot{M}/M)^{(1/4)}$ values shown by
Kishimoto's sample of Quasars. However, the power to $1/4$ clearly
introduces a weak dependency on $\dot{M}/M$, and given the
uncertainties involved in these measurements it is still premature to
draw firm conclusions. In fact, NGC~4395 (for which the detected
intra-day near-IR variations also argue in favor of J and H-band disc
emission; Minezaki et al., 2006), has a value of $(\dot{M}/M)^{(1/4)}$
well consistent with the other NGC galaxies. Hence, even though Table
5 could be telling us that MR~2251-178 hosts the coolest accretion
disc it remains to be tested further whether the combination of these
two fundamental parameters, $M$ and $\dot{M}$, is sufficient to fully
determine the temperature distribution in thin accretion discs. As
already shown in this work the geometry of the disc (e.g., the
presence of flaring, humps, or tapering of the disc) might also play
a significant role.

This finding could have important implications for the analysis of other
sources with small $(\dot{M}/M)^{(1/4)}$ ratios where the near-IR emission is
normally interpreted as a dust emission while it might correspond to emission
from the accretion disc.

\section{Conclusions}

We have obtained high quality near-IR light curves for two nearby AGN:
NGC~3783 and MR~2251-178, which are characterised by very different
values of their black hole mass. The light curves track variations on
time-scales from a few days up to three years for NGC~3783 and four
years for MR~2251-178. From the analysis of these data, together with
the results already presented in Ar\'evalo et al.\ (2008, 2009) we
find the following:

\begin{enumerate}

\item The Near-IR light curves for both our sources, NGC~3783 and MR~2251-178,
  show a strong correlation with the optical light curves.
\item For NGC~3783 the optical and near-IR cross correlation analysis
  suggests that two emitting regions are present in the J-band
  emission, one consistent with a short lag and another found much
  further out. We identify these regions as the accretion disc and a
  dusty torus. The H and K lags ($\sim 70$ days) are consistent with
  the presence of dust in a torus located at the sublimation radius.
\item For MR~2251-178 the lags measured between all near-IR and optical
  wavelengths are consistent with lags of $\sim 0$ days. This suggests that
  the variable near-IR emission arises from the accretion disc in this source.
\item From the cross correlation analysis we find no direct evidence
  for the presence of a dusty torus in MR~2251-178 as expected from
  the delays predicted by the Suganuma et al.\ (2006) relation. We
  checked that the emission is not due to the presence of beaming
  emission from a radio jet.
\item Flux-flux plots for both sources are consistent with the
  findings from the cross correlation results. The linear relations
  observed between the optical and near-IR bands in MR~2251-178
  confirm that the variable near-IR emission originates in the
  accretion disc. However, some near-IR excess seen at low optical
  fluxes might suggest the presence of a dusty torus.
\item We determined a satisfactory SED representation for
  NGC~3783 adopting a model of an $\alpha$-disc illuminated by a
  central X-ray source plus a dusty torus at a distance consistent
  with the dust sublimation radius. The model is able to reproduce the
  optical and near-IR spectral energy distribution as well as other
  timing constraints. The modelling of MR~2271-178 is not as adequate
  and requires the presence of a dusty torus.
\item We tentatively interpret the differences in the origin of the
  near-IR emission in our sources, as well as those reported by
  Suganuma et al.\ (2006) and Kishimoto et al.\ (2009), as the result
  of the location of the colder regions in the accretion disc with
  respect to the central illuminating X-ray sources, as originally
  suggested by Uttley et al.\ (2003).

\end{enumerate}

\section*{Acknowledgments}

We thank the anonymous referee for helpful comments. PL gratefully
acknowledges support by Fondecyt Proyect 1080603 and Fondap Project 1501003.

\end{document}